\documentclass[11pt]{amsart}
\usepackage{graphicx}
\usepackage{float}
\usepackage{amssymb,amsmath,amsthm}
\usepackage{epstopdf}
\usepackage{color}
\usepackage{mdframed}
\usepackage{empheq}
\usepackage{bm}
\usepackage{hyperref}
\usepackage{mathabx} 
\usepackage{enumitem}
\usepackage{xcolor}
\usepackage[linesnumbered,ruled,vlined]{algorithm2e}
\usepackage{subcaption}

\newtheorem{theorem}{Theorem}

\newtheorem{remark}[theorem]{Remark}

\makeatletter
\renewcommand\l@paragraph[2]{}
\renewcommand\l@subparagraph[2]{}
\makeatother
\title[Glasso]{Joint Stochastic Model for Electric Load, Solar and Wind Power at Asset Level and Monte Carlo Scenario Generation}
\author{Ren\'e Carmona \& Xinshuo Yang}
\address{Department of Operations Research \& Financial Engineering\\
Princeton University\\
Princeton, NJ 08544, USA}
\date{February 20, 2021}                   

\begin{document}
\maketitle
\hfill

\begin{abstract}

For the purpose of Monte Carlo scenario generation, we propose a graphical model for the joint distribution of wind power and electricity demand in a given region. To conform with the practice in the electric power industry, we assume that point forecasts are provided exogenously, and concentrate on the modeling of the deviations from these forecasts instead of modeling the actual quantities of interest. We find that the marginal distributions of these deviations can have heavy tails, feature which we need to handle before fitting a graphical Gaussian model to the data. We estimate covariance and precision matrices using an extension of the graphical LASSO  procedure which allows us to identify temporal and geographical (conditional) dependencies in the form of separate dependence graphs.
We implement our algorithm on data made available by NREL, and we confirm that the dependencies identified by the algorithm are consistent with the relative locations of the production assets and the geographical load zones over which the data were collected.  
\end{abstract}


\section{\textbf{Introduction}}
\label{se:introduction}

Modern operation of the electric grid has become highly technical but despite clear evidence of the stochasticity of electricity demand the major unit commitment and economic dispatch computer algorithms remained intrinsically deterministic. This is the more problematic given the increased penetration of renewable energy production sources like solar and wind whose uncertainly and volatility have been universally documented. For this reason, it became 
commonplace to have engineers rely on probabilistic forecasts of wind, solar and even loads in lieu of deterministic point forecasts of these quantities.
Even though these new classes of forecasts provide a form of quantification of the variability of the variables, and hence of the reliability of the point forecasts, these forecasts remain very limited in the sense that they have no ability to capture temporal and spatial dependencies between the random quantities they try to predict. Still they have been the object of intense interest during the last decades and the sources of a very large number of publications. For the methodologies behind these probabilistic forecasts we refer to the following recent surveys: \cite{HongFan} for load, \cite{ZhangWangWang} for wind power, and \cite{SolarForecasting} for solar power generation.
While the application domain is the same, our approach differs significantly from those papers. Indeed, not only                                                                                                                                                                                                                                                                                                                                                                                                                                                                                                                                                                                                                                       do we  not produce forecasts, but we use existing forecasts and we model the deviations of the actual values from their forecasts. This choice is based on the fact that many users have their own forecasts which they do not want to part from, whether they purchase them or they develop them in house,  

Beyond the interest in the production of forecasts, a large number of studies were devoted to the development of algorithms to generate statistical scenarios from probabilistic forecasts. For wind power, we refer to the fundamental paper \cite{Pinson_et_al_1} and the references therein, and to \cite{WatsonWets} for a recent \emph{state of the art} on the subject with a clear goal of providing inputs for unit commitment and economic dispatch platforms. Evaluation of the quality of scenarios produced this way was a natural development. See for example \cite{PinsonGirard} for the analysis of wind scenarios.

Unfortunately, spatial and temporal correlations were very rarely included in the simulation models, and when it was done, it was often for one single variable only. See \cite{Pinson_et_al_2} for an example dealing with wind or \cite{TastuPinsonMadsen} for a spatio-temporal model for wind power. One of our motivations, and possibly the main thrust of our paper, is to propose algorithms to remedy this shortcoming by allowing temporal and spacial dependencies to be captured for load, wind power and solar power simultaneously, even at the most granular level of the individual power plants. Modeling the random behavior of the future values of a variable, say wind, is commonly done on the basis of existing forecasts obtained exogenously. See for example \cite{wind_power}. This trend has become a standard in the industry, and we shall follow this time honored convention in the present paper.

\subsection{Contributions of the paper.}
This paper is a contribution to the literature on the modeling of the stochastic behavior of load, wind and solar power at their most granular level available, and the development of simulation algorithms capable to deliver Monte Carlo scenarios for their future time evolutions. 
While often touted in the published literature,  the analysis of fully correlated models is very rarely pushed beyond the level of ideas and discussions, and even more rarely implemented by means of simulation algorithms. See nevertheless the inclusion of time correlation for wind power in \cite{Morales_et_al,TastuPinsonMadsen}.  We do not know of any prior work involving joint spatio-temporal models for load, wind and solar.

Before delving into the gory details of our modeling approach and the inner working of our simulation engines, we want to emphasize that for all the time series of interest, we work with the deviations from forecasts. This is a time honored approach which is widely accepted in the engineering community, and for this reason, which we follow. The down side is that the historical data needed to fit our model needs to include actual measurements of load, wind and solar power as well as their respective forecasts. Our models and their implementations as simulation engines are agnostic to the source of the forecasts. However, the quality of the latter does impact the goodness of fit of our models as well as the predictive power of the Monte Carlo simulator. 

\vskip 1pt
We now describe in broad brushstrokes the several prongs of our approach, in no small part, to emphasize the originality of our contribution. First, we recognize that the marginal distributions of the hourly zonal load variables have heavy tails. So our first step is to check for the presence of heavy tail in marginal hourly measurements, and when found, fit a Generalized Pareto Distribution (GPD) to the marginals. We use them to process the data and give them Gaussian marginals. We explain and illustrate the pros and cons of this step with simulation illustrations on days for which the forecasts overestimate the actual future values and days for which they underestimate the actuals. In some respect, our approach is reminiscent of the Gaussian copula approach as advocated for example in \cite{Pinson_et_al_1} and \cite{TastuPinsonMadsen}. However, our approach differs in significant ways from theirs. First because we fit heavy tail distributions to the marginals, and then we use non-parametric estimation techniques based on the LASSO $L^1$ regularization to estimate large covariance matrix instead of the parametric forms used in \cite{TastuPinsonMadsen}.
Another major feature of our model and its implementations is to be usable when the number of historical observations used to fit the model is small compared to the dimension of the state variables we model and simulate. This issue is often encountered in modern high dimensional statistics. One of the major issues with this feature of the data is that it implies a systematic singularity of the empirical covariance matrix estimates. In order to overcome this shortcoming, we use an approach based on graphical Gaussian models estimated by a form of the graphical lasso algorithm {\tt glasso}. In this way, we obtain reasonable sparse proxies for the precision matrices, and work with Gaussian models which can be simulated. As a further step to reduce the number of degrees of freedom we use the {\tt gemini} algorithm  introduced in \cite{Zhou}, to search for precision matrices in the form of tensor products, allowing us  to disentangle the contributions of the spatial and temporal components to the correlation structures.

\vskip 4pt
The model introduced in this paper is built from the ground up. We first introduce marginal models separately for the three features: load, wind power and solar power. Historical data for load actuals and forecasts being only available at the zone level, possible correlations between load and wind on one hand, and between load and solar on the other, need to be detected and implemented at the zonal level, even if eventually, wind and solar power need to be modeled and simulated at the individual asset level.

The marginal model for load at the zone level is inspired by our earlier work \cite{CarmonaYang} on the ERCOT data publicly available from the Texas ISO web site.  While the load data used in the present paper is very different from the data used in \cite{CarmonaYang}, load data is still at the zone level, and even though the zones are different in their number and coverage, the modeling approach used in the present paper is similar to the one of our previous work. The stochastic models we developed for load, and wind and solar power are the fulcrum for a Monte Carlo simulation engine providing scenarios on demand. Given a date and a set of historical observations (including forecasts) prior to that date, the simulation engine produces as many Monte Carlo scenarios as requested for the next $24$ hourly values of load, wind and solar,    

The simulation engine outputs scenarios in the  format predicated by the format of the historical data used to fit the models.
A first version of the simulation engine was implemented in a Python package by Michal Grzadkowski and made available with extensive README files and 
examples which can be run from Jupyter Notebooks. It can be found at \cite{PGscen}.

\subsection{Organization of the paper.} 
The paper is organized as follows: in Section \ref{se:data} we describe the data used for the analysis. The following Section \ref{se:load} is devoted to the analysis of the aggregate demand for electric power at the zone level.  
The fitted model is the basis for the Monte Carlo simulation engine whose performance is demonstrated by examples of batches of scenarios.  Our asset level model is developed and fitted to the wind power generation historical data  in Section \ref{se:wind}.
It is shown there that, when aggregated at the zone level, no significant dependence with load is detected. The situation is very different for solar power.
The first part of Section \ref{se:solar} is devoted to the development of a marginal model for the $24$ hours of solar power production for each of the $222$ solar stations for which we have historical data. The second half of the section is devoted to the aggregation of the solar production at the level of the load zones and to the quantification of the dependencies between load and solar at these levels.
Load and solar generation are modeled together in Section \ref{se:altogether}. The main take-away from this section is that, when designing Monte Carlo simulation engines from the model, we first simulate load and solar power at the aggregate zone level and then, from the zone solar production scenarios generated jointly with load, we simulate $24$-dimensional solar generation for each asset, conditioned by the value of the vector of aggregate production at the zone level already simulated.

\vskip 2pt
We conclude with a short recall of the main contributions of the paper in Section \ref{se:conclusion}.

\section{\textbf{Problem Set-Up and Description of the Data}}
\label{se:data}

\subsection{Case Study Data}
\label{sub:data}
For the purpose of illustration and demonstration, we use the first of the data sets assembled  by NREL for the Department of Energy program ARPA-E PERFORM, \cite{NREL}. The data covers the Texas region. This synthetic data set contains time series for electricity demand at the regional level, and actuals and forecasts of wind and solar power generation at the asset level. Several time frequencies are provided, but for the sake of definiteness, we shall limit ourselves to hourly data. The geographical resolution is the highest in the sense that data is provided at the site-level. 

\vskip 4pt
The \textbf{solar} data comprises $226$ sites (including $22$ existing and $204$ proposed solar sites) in the Texas region. The solar power forecasts are provided for $2018$ as probabilistic forecasts in the form of $1$ to $99$ percentiles. These probabilistic forecasts are generated using a Bayesian Model Averaging (BMA) \cite{NREL} procedure to the ensembles of $51$ deterministic forecasts provided by the European Centre for Medium-Range Weather Forecasts (ECMWF). However,  since the ECMWF forecasts included in the NREL data set are for $2017$ as well as $2018$,
in order to maximize the size of historical data to which we fit our model, we chose to ignore the NREL probabilistic forecasts and use instead, a plain average of the $51$ ECMWF deterministic forecasts as a point forecast.
In summary, the day-ahead forecasts we use have a $12$-hour-ahead lead time, a $24$-hour horizon, an hourly resolution, and a daily update rate.
For each day, at $12:00$ (issue time), we have a set of hourly forecasts for the next day starting from $00:00$ ending at $23:00$ (forecast time).
For the actual solar power, the data is given with a $5$-min resolution for $2017$ and $2018$, so we computed an average for each hour to obtain hourly solar power.
 
\vskip 4pt
The \textbf{wind} data comprises $264$ sites (including $125$ existing and $139$ proposed sites) in the Texas region. 
The wind day-ahead forecasts are provided by NREL in the same format as the solar data – probabilistic forecasts for $2018$ and $51$ ECMWF forecasts for $2017$ and $2018$. However, the ECMWF forecasts are highly biased and point forecasts obtained by plain averaging yield much larger deviations. For this reason, we prefer to use the median of the probabilistic forecasts as a point forecast. However, in order to demonstrate the impact of the quality (or lack thereof) of the forecasts on the performance of the output scenario generation, we also use the plain average of the $51$ ECMWF forecasts. Note that for the median of the probabilistic forecast,  the available historical data to fit and test the wind model is limited to $2018$ only, while for the average of the ECMWF forecast, the available historical data covers $2017$ and $2018$.  For the actual wind power, the data is given with a $5$-min resolution for $2017$ and $2018$. Again, we computed an average for each hour to obtain hourly wind power data.

\vskip 4pt
The NREL data set also contains \textbf{load} data at the zone level. In fact, the load actual and forecast data were collected from the ERCOT official website at hourly resolution for $8$ load zones.
The load forecasts were obtained from ERCOT's "Seven-Day Load Forecast by Weather Zone", see \cite{ERCOT}, by selecting the data issue and forecast times to match those of the solar day-ahead forecasts.

\begin{remark}
While the examples given in this paper are based on the NREL synthetic data set described above, we also implemented our model on the IEEE's Reliability Test System (RTS) whose update is presented in \cite{RTS}. The interested reader can obtain simulation results for RTS by running the examples provided in the Jupyter Notebooks which can be found on the GitHub repository of the simulation platform PGscen, \cite{PGscen}.
\end{remark}

\section{\textbf{Marginal Model for Electricity Loads}}
\label{se:load}

We first tackle the challenge of the design of the Monte Carlo scenario simulation engine for the loads over the region.

\subsection{\textbf{Strategy}}
As explained in the introduction, we model the deviations (which are often called errors in the literature on the subject) defined as
\begin{center} 
\emph{load deviations} = \emph{load actuals} – \emph{load forecasts}.
\end{center}
Recall that the historical data  we secured covers the period 2017-01-01 to 2018-12-31. For each day $d$ and for each load zone 
$$
z\in\{\text{Coast}, \text{East}, \text{Far West}, \text{North}, \text{North Central},\text{South},\text{South Central},\text{West}\}, 
$$
we read the vector of the $N_{lag}=24$ load point forecasts from the load forecast data set, and for each lag $\ell\in\{0,1,\cdots,N_{lag}-1\}$, we subtract the point forecast from the actual load found in the load data set.
This gives what we call the load deviation. The data set of all deviations so constructed has the same structure as the data set of load forecasts. 

\vskip 12pt\noindent
1)  For each load zone $z$ and for each lag $\ell\in\{0,1,\cdots,N_{lag}-1\}$ the load deviations form a daily time series of length $N$,  the number of days prior to the day we generate scenarios for. The first step of our analysis is to analyze the marginal distributions of these $N^L_{zone}\times N_{lag}=8\times 24=192$ stationary univariate time series.

\begin{figure}[H]
\centerline{
\includegraphics[width=4cm,height=4cm]{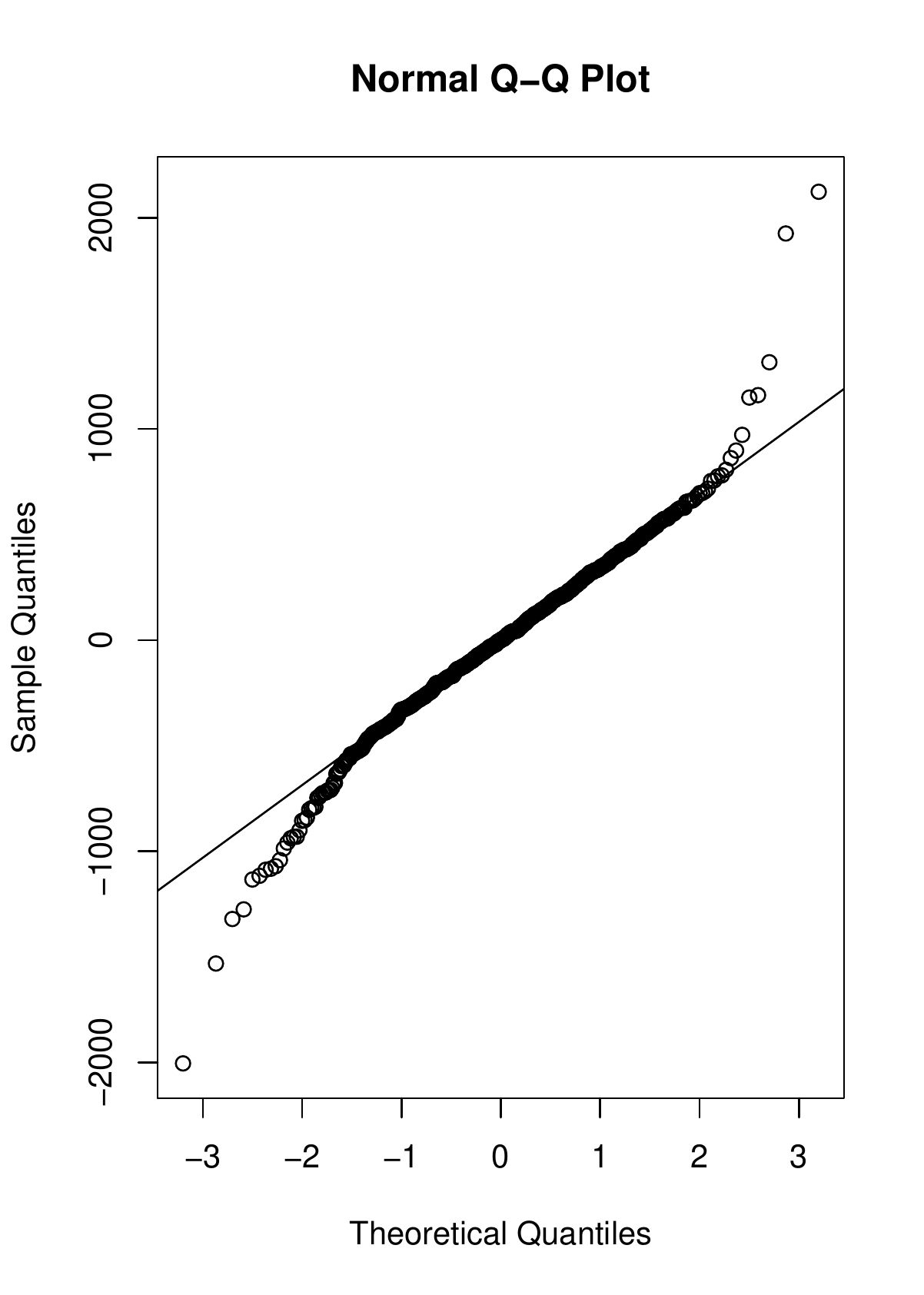}
\hskip 2pt
\includegraphics[width=4cm,height=4cm]{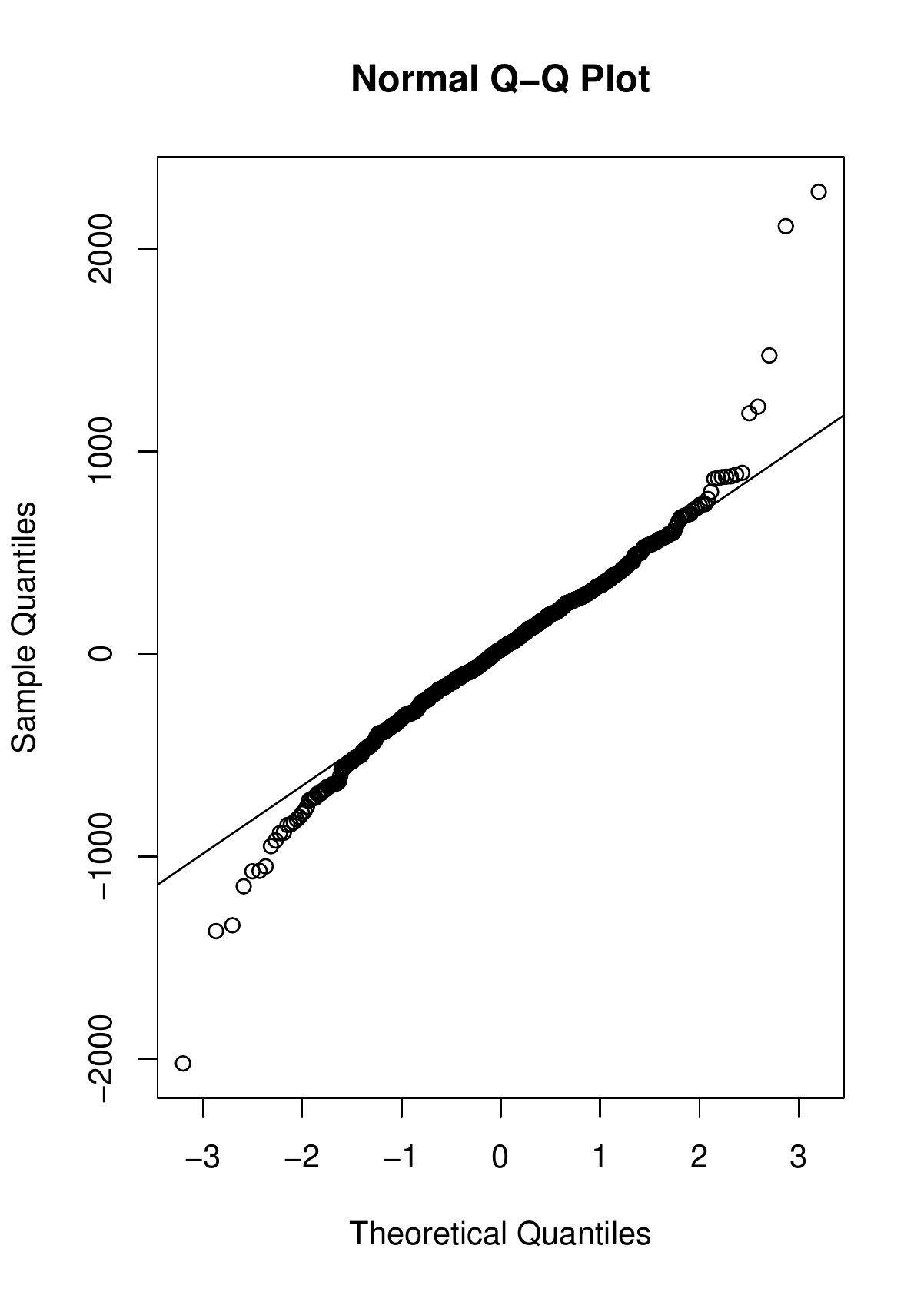}
\hskip 2pt
\includegraphics[width=4cm,height=4cm]{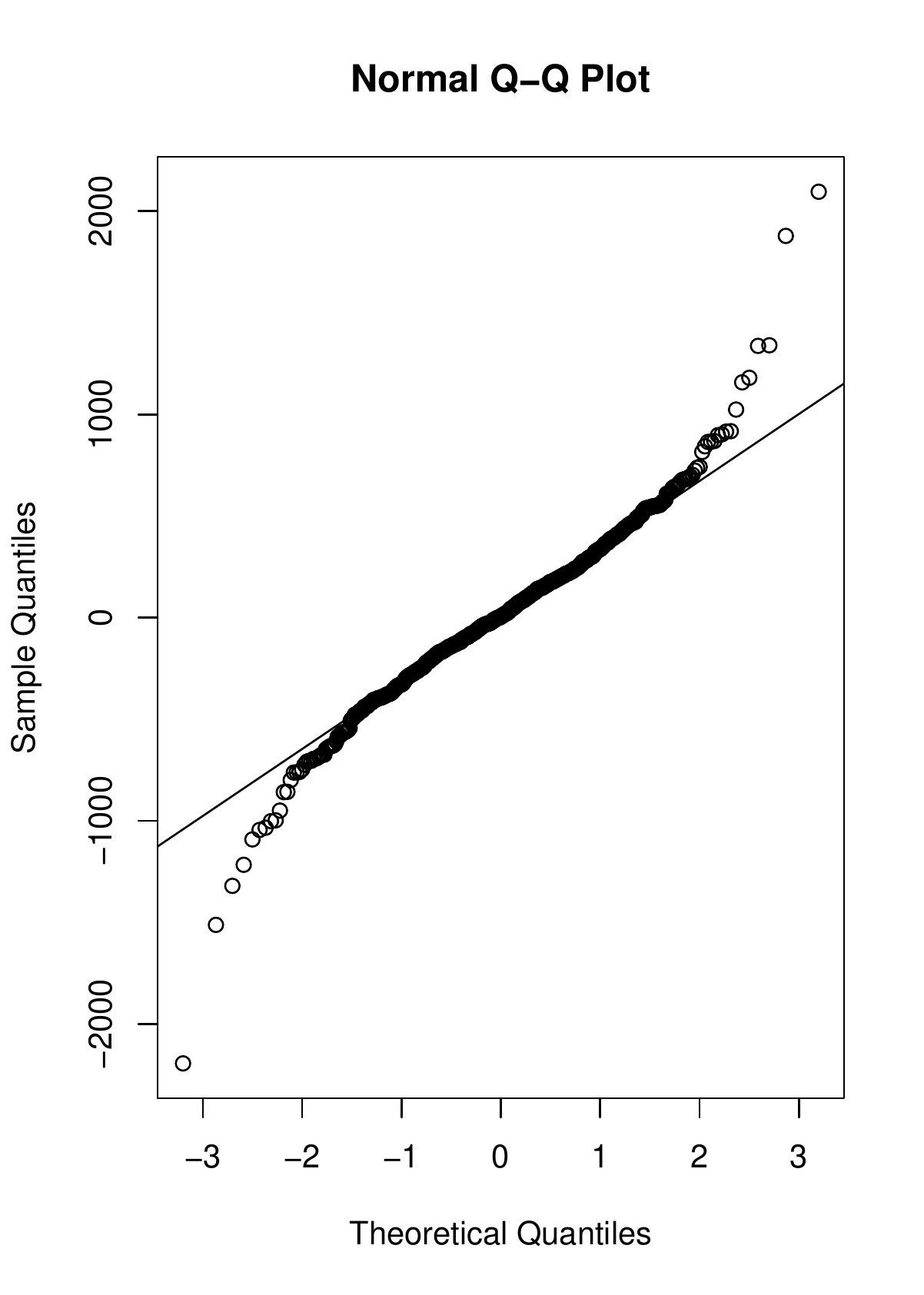}
\hskip 2pt
\includegraphics[width=4cm,height=4cm]{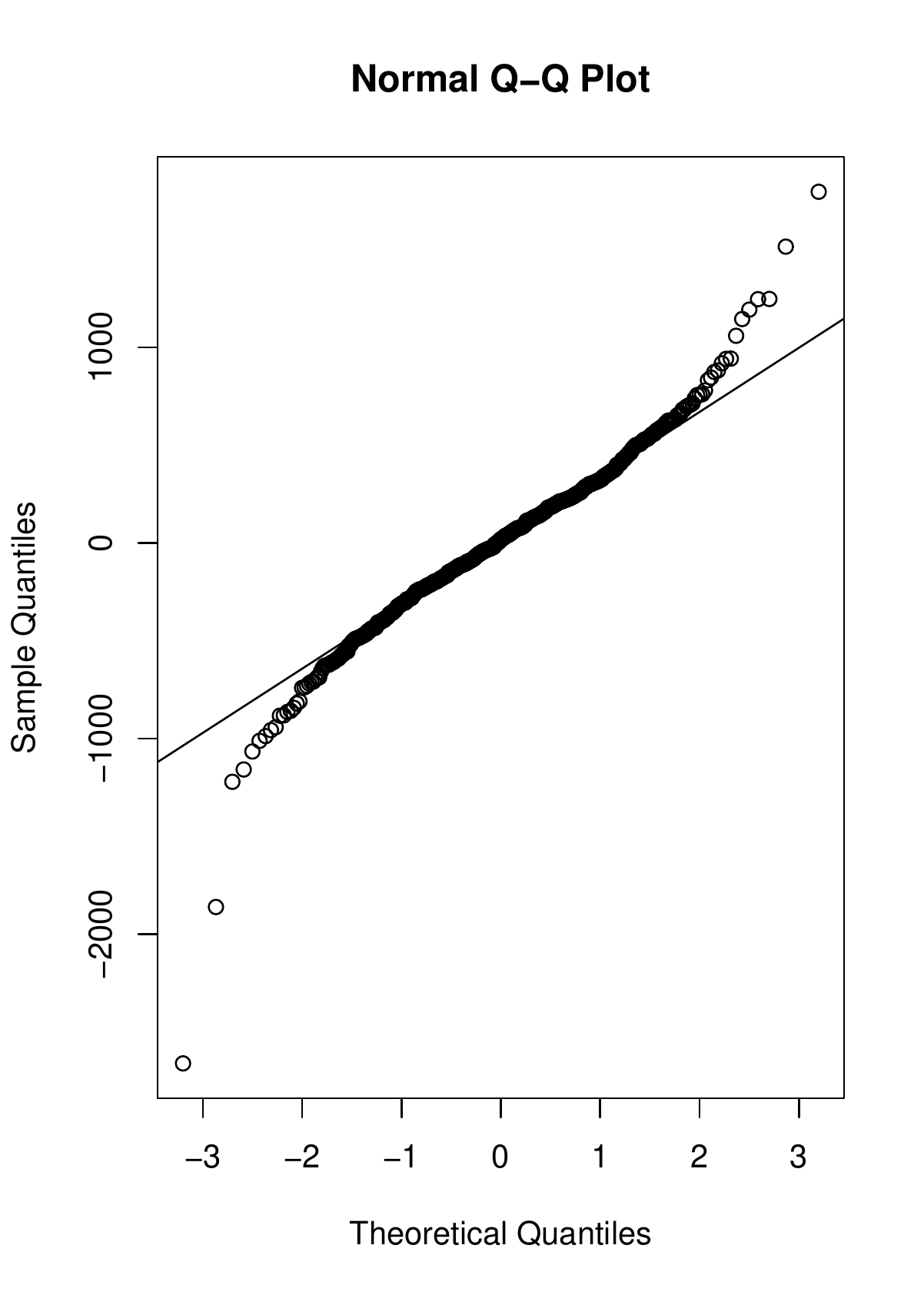}
}
\caption{Normal Q-Q plots for load deviation remainders in the $z=\text{Coast}$ region and lags $\ell=0, 1, 2$ and $\ell=3$ hours ahead.}
\label{fi:loads_qqnorms}
\end{figure}

\vskip 4pt\noindent
2) For each load zone $z$ and for each lag $\ell$ we produce a Q-Q plot of the empirical quantiles of the load deviations against the theoretical quantiles of the standard Gaussian distribution $N(0,1)$. Figure \ref{fi:loads_qqnorms} give a sample of $4$ of these Q-Q plots. They are typical of what we found throughout. This is clear evidence that the marginal distributions of the load deviations have \emph{heavy tails}, justifying our next step. 

\vskip 4pt\noindent
3) For each load zone $z$ and for each lag $\ell$ we fit a Generalized Pareto Distribution (GPD), see for example \cite[Chapter 2]{Carmona_SAFD}, to the  load deviations. We denote such a distribution as $G^L_{z,\ell}$ for the sake of later reference. Throughout this work, in order to manipulate GPD distributions, we use functions of the {\tt R} library {\tt Rsafd} \cite{rsafd} developed for the analysis of the examples and problem sets of the book \cite{Carmona_SAFD}, which we wrapped into a Python package.

\begin{figure}[H]
\centerline{
\includegraphics[width=4cm,height=4cm]{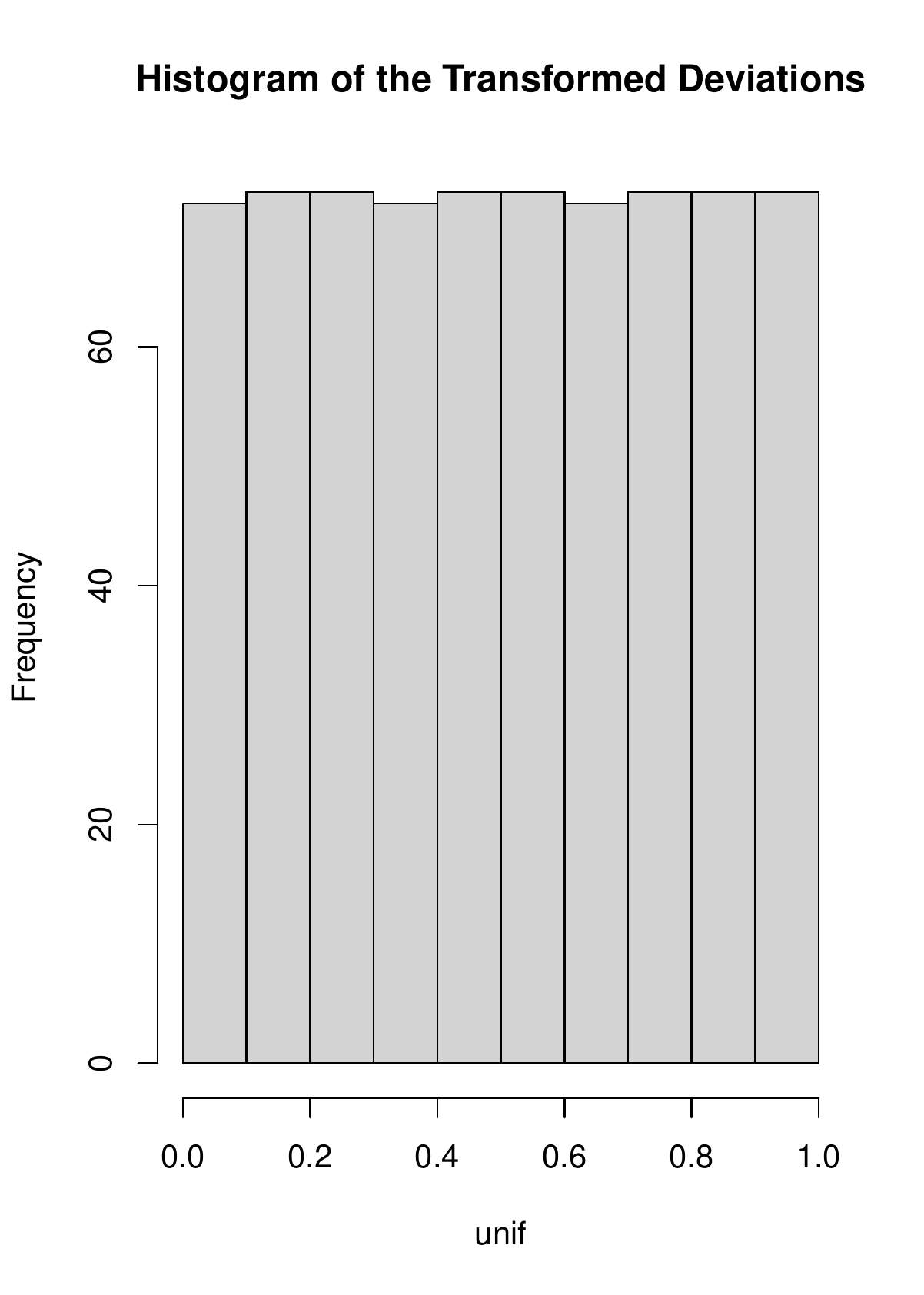}
\hskip 6pt
\includegraphics[width=4cm,height=4cm]{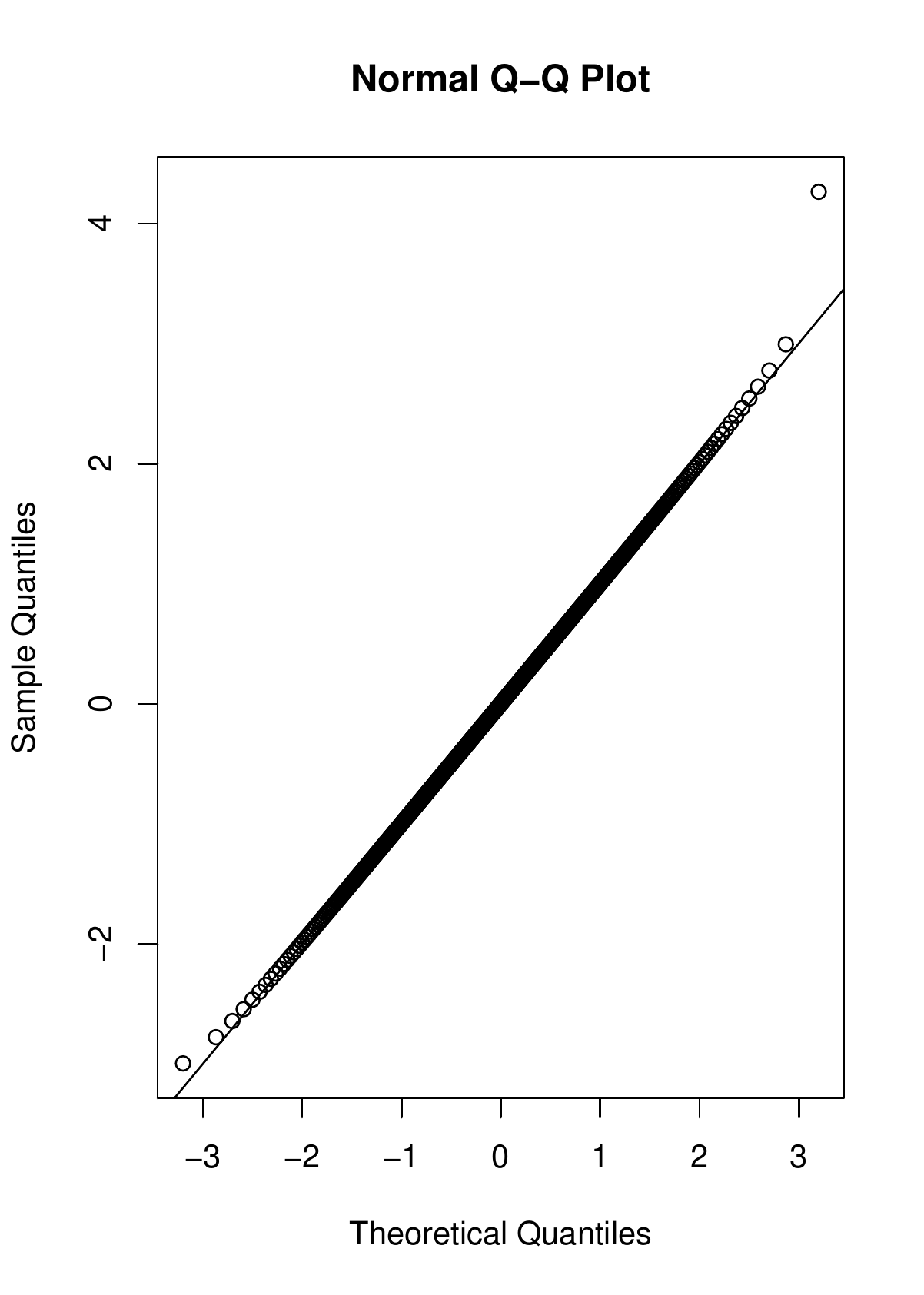}
}
\caption{Transformations of the load deviations in the Coast region $4$ hours ahead.}
\label{fi:load_tranform}
\end{figure}

\vskip 4pt\noindent
4) For each load zone $z$ and for each lag $\ell$ 
we transform the load deviation time series into a uniform time series by computing the cumulative distribution function of the GPD just fitted on each of the entries of the time series. If the fitted GPD is a good fit, we should expect the marginal distribution of the new time series to be uniform. The histogram on the left pane of Figure \ref{fi:load_tranform}  is a testament to this claim.

\vskip 4pt\noindent
5) For each load zone $z$ and for each lag $\ell$ we transform the uniform time series just obtained into a time series with a standard Gaussian distribution $N(0,1)$. In order to do that, we just compute the quantile function of the standard Gaussian distribution on each of the entries of the uniform time series. The normal Q-Q plot on the right pane of Figure \ref{fi:load_tranform} gives a graphical evidence of the fact that the marginal distribution of the resulting time series is indeed a standard Gaussian distribution.

\vskip 4pt\noindent
6) At this stage of the analysis, for each day $d$, we have a $N^L_{zone}\times N_{lag}=8\times 24=192$ dimensional numerical vector, each component of this vector having a standard Gaussian distribution. While we do not know if this $192$-dimensional vector is actually \emph{jointly Gaussian}, we shall behave as if that was. Given the high dimensionality of the vectors when compared to the small amount of historical data available to estimate the covariance structure of such a Gaussian vector, we expect that any empirical estimate of the covariance matrix will be degenerate and hence not invertible. So in order to estimate the precision matrix capturing the dependencies between the various components of the vector, we use a regularization procedure and we proceed to fit a \emph{Gaussian Graphical Models} using {\tt LASSO}. The interested reader will find a readable introduction to Gaussian graphical models, including LASSO estimation in \cite[Chapter 11]{Wainwright}. Here we could use the {\tt R} package {\tt glasso} to fit the graphical model to our modified load data. However, we believe that we would miss the special structure of our $192$ dimensional vector which has a $8$-dimensional spatial component and a $24$-dimensional temporal component. Taking one more step to reduce the mismatch between the high dimension of the vector and the short historical data, and relying on the intuition developed for classical statistical analysis of variance with multiple factors, we would like to think of the $192\times 192$ covariance and precision matrices as the tensor product (also known as Kronecker products) of a $8\times 8$ covariance and precision matrices capturing the dependencies between the spatial components, and a $24\times 24$ covariance and precision matrices for the temporal component. Searching for a precision matrix and a covariance matrix in this form dramatically reduces the number of parameters to be estimated. In the statistical literature, these special covariance matrices are called \emph{separable}. Testing for the presence of such a special structure is sometime possible. See for example \cite{separability_test_1,separability_test_2} and the more recent paper \cite{separability_test_3}, even though none of these tests have enough power in our case because of the lack of historical data.  In any case, we shall work under this assumption here, and we shall use the variation on the {\tt glasso} algorithm called {\tt GEMINI} which was proposed in \cite{Zhou} to take advantage of the separable structure.

\begin{figure}[h]
\centerline{
\includegraphics[width=6cm,height=6cm]{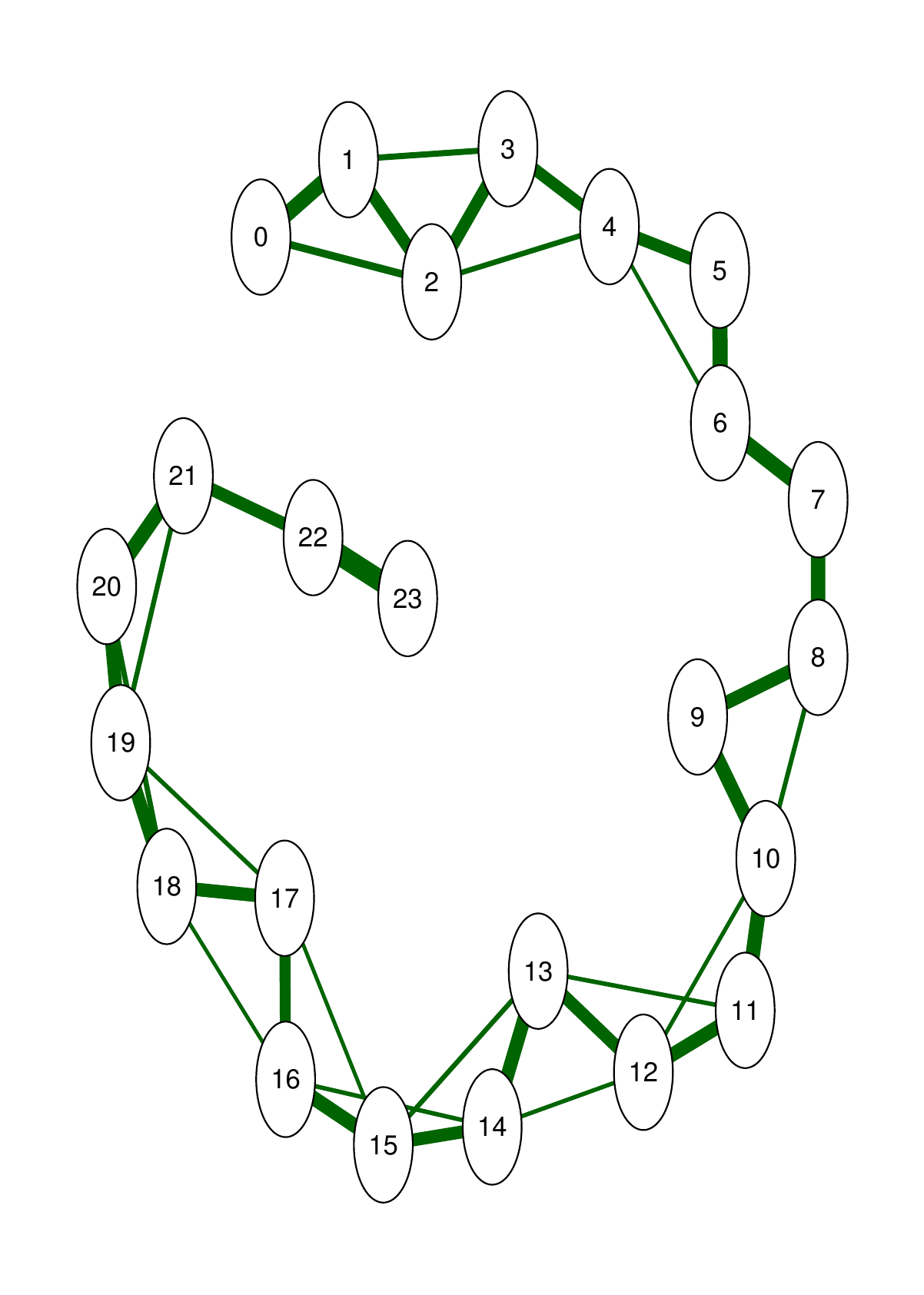}
\hskip -2pt
\includegraphics[width=6cm,height=6cm]{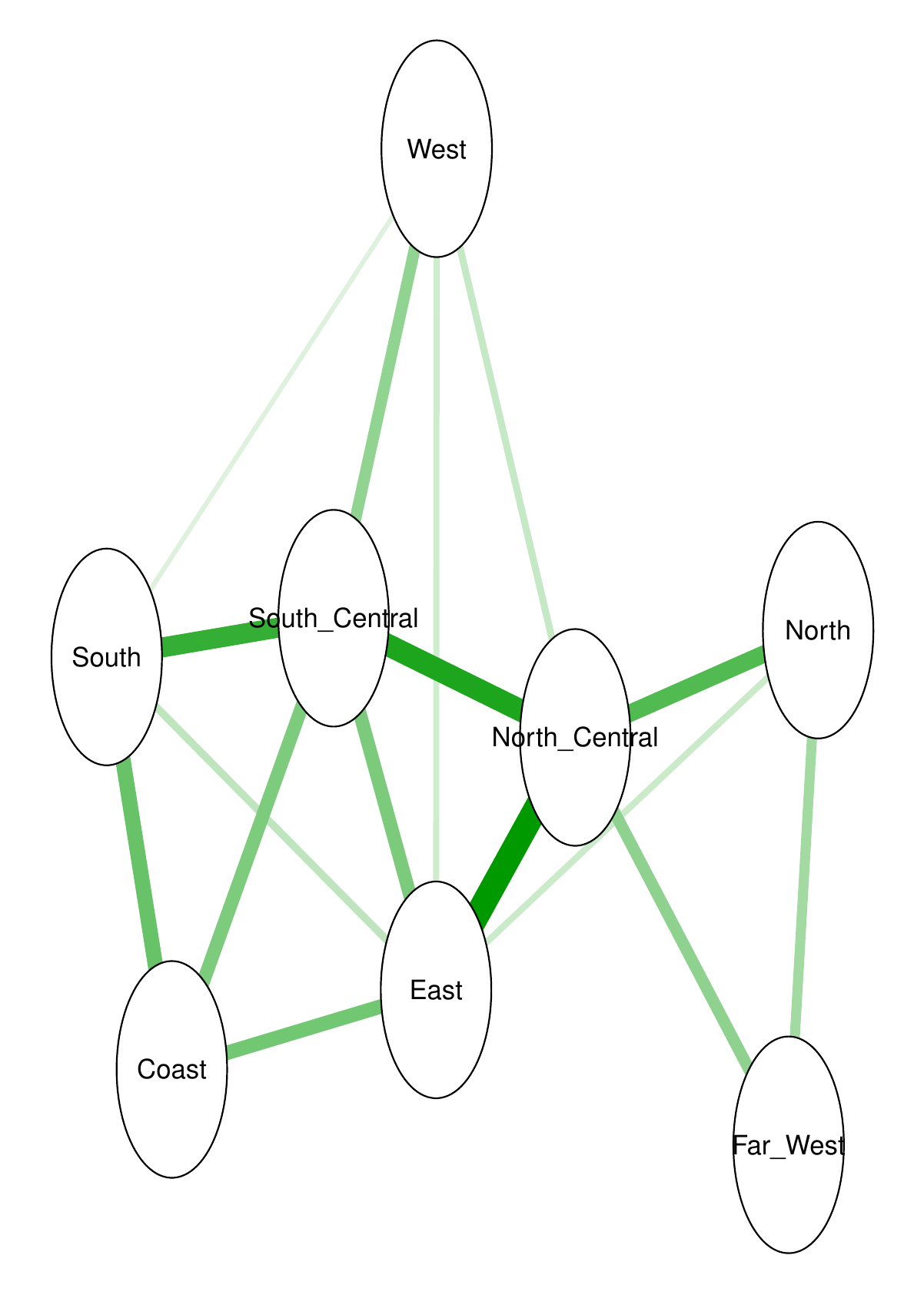}
\hskip -2pt
\includegraphics[width=6cm,height=8cm]{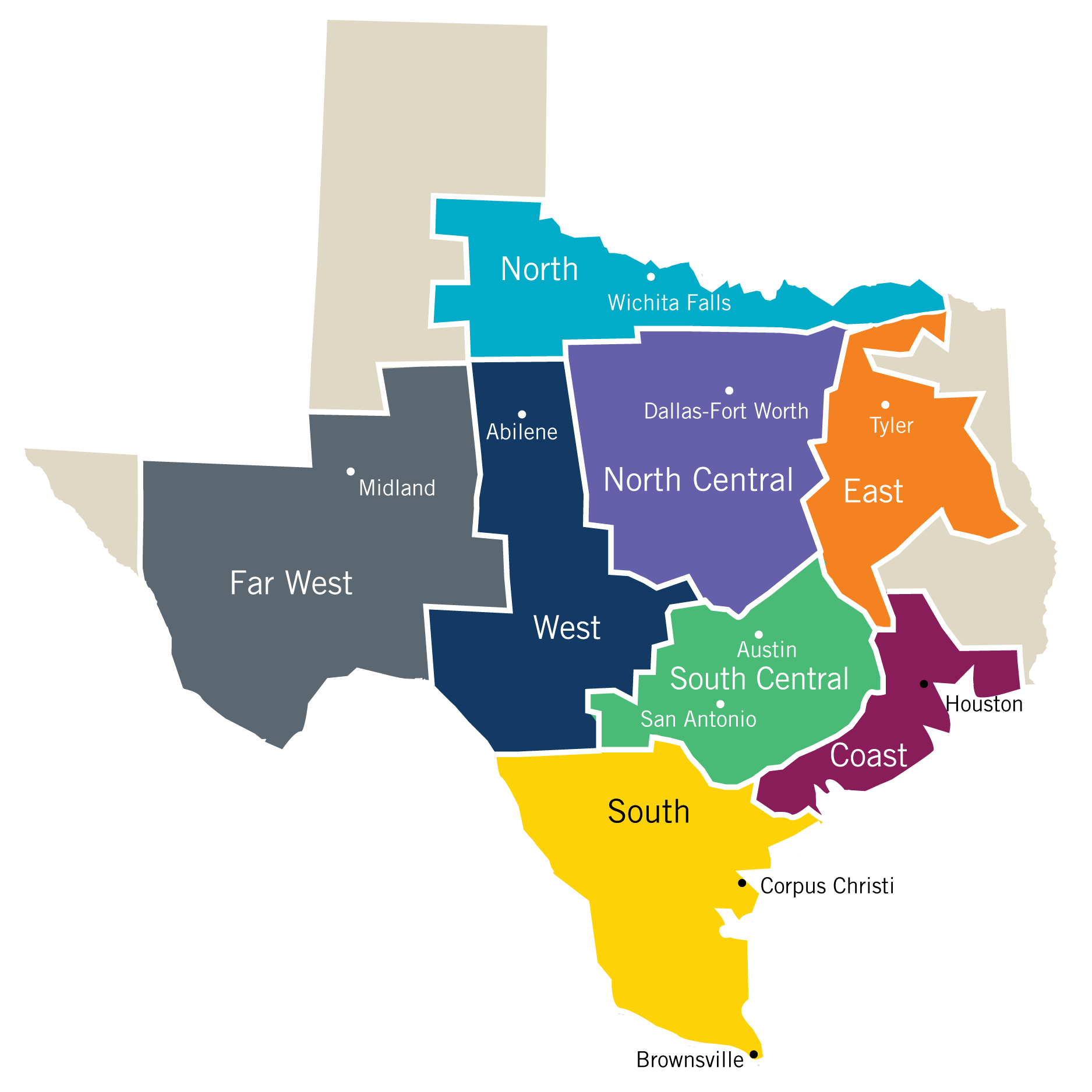}
}
\caption{Temporal component of the conditional correlation (left), spatial component of the conditional correlation (middle), geographical map of the 8 load zones in ERCOT.}
\label{fi:load_glasso}
\end{figure}

\vskip 4pt
Figure \ref{fi:load_glasso} shows the graph of the conditional temporal dependence in the left pane, as well as the spatial conditional dependence component provided by the {\tt GEMINI} algorithm. 
\vskip 12pt

The graph of the temporal dependencies comprises two types of links. The thick links form a perfectly ordered linear chain. It points clearly at some form of Markovian structure. However, the thinner links show a residual dependence for hours two lags apart, which is quite intuitive. The edges, as well as their strengths in the graph of the spatial dependencies form a pattern consistent with the contiguity of the load regions as shown in the right pane of Figure \ref{fi:load_glasso}.

\vskip 4pt\noindent
7)  The next step of the analysis is to generation of Monte Carlo samples from the model. Given the covariance matrix estimated above, we can easily generate as many Monte Carlo samples from the $192$-dimensional mean zero Gaussian distribution with this covariance matrix. Computing the cumulative distribution function of the standard Gaussian distribution on each of the entries provides sample of $192$-dimensional random vectors with uniform marginals. Each of the $192$ components corresponds to a specific zone $z$ and a specific lag $\ell$ so if we now compute the quantile function of the GPD distribution $G^L_{z,\ell}$ fitted to the original load deviations for zone $z$ and lag $\ell$, we obtain samples of a $192$ dimensional random vector with marginal distributions originally fitted to the load deviation time series. At this stage, it is important to emphasize that these $192$ components depend upon each other in the way dictated by the covariance structure found in the estimate of the graphical Gaussian model. The astute reader will have certainly noticed that we are actually using a $192$-dimensional copula to build the dependencies between the $G^L_{z,\ell}$ marginals of this $192$-dimensional vector.

At this stage, it is very easy to complete the simulation cycle by adding back the forecasts, in order to end up with Monte Carlo scenarios of the actual loads.

\begin{figure}[H]
\centerline{
\includegraphics[width=16cm,height=8cm]{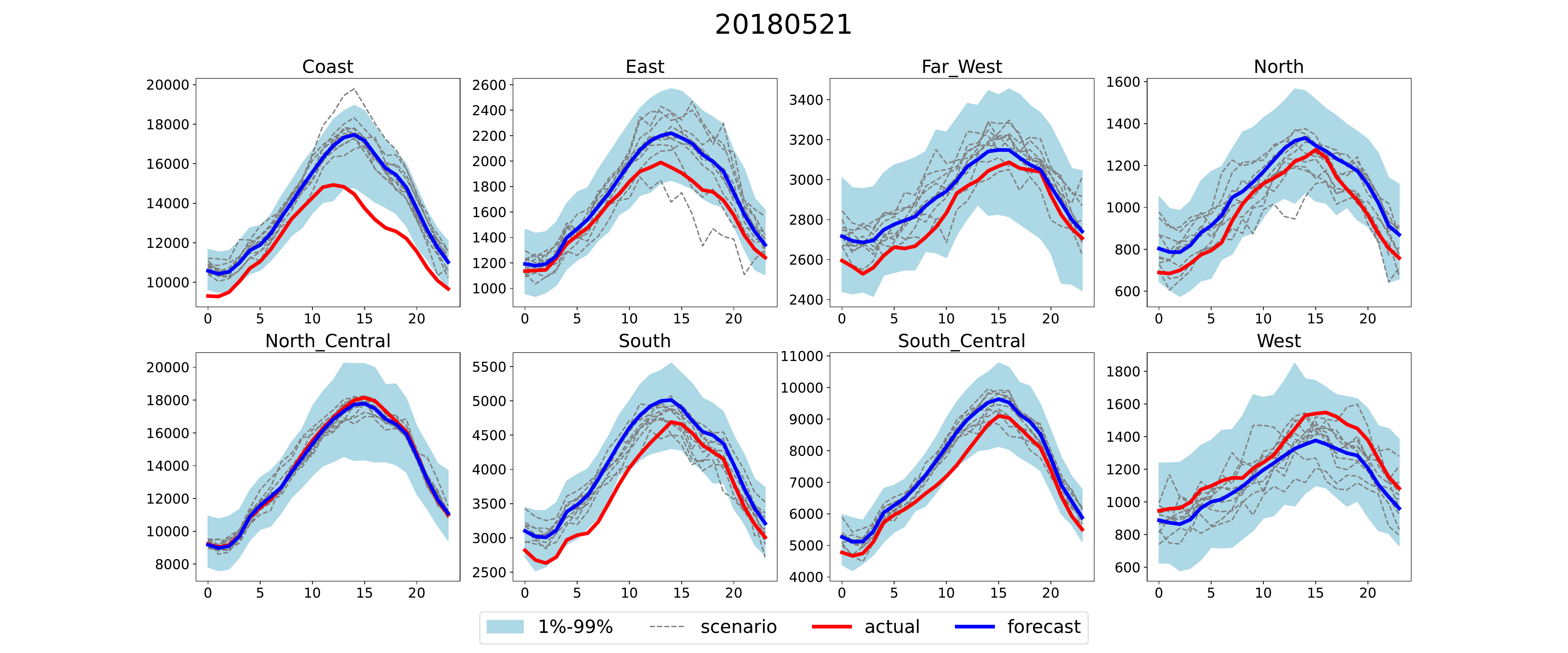}
}
\caption{$1000$ Monte Carlo scenarios  generated for the next $24$ hours for May 21, 2018. The red line gives the actual load demand observed over these $24$ hours while the blue line shows the point forecasts. The dashed lines are $10$ randomly selected scenarios. The light blue bands are the traces left by the Monte Carlo scenarios after we remove the smaller $1\%$ and the larger $1\%$ of the bunch.}
\label{fi:load_scenarios}
\end{figure}

Figure \ref{fi:load_scenarios} shows what the scenarios look like vis-a-vis the actual and forecasts loads for May 21, 2018.

\begin{remark}
Notice that we assumed some form of stationarity of the load time series as we worked under the assumption that the load deviations have a single maarginal distribution, irrespective of the day of the year (whether $d$ is in the summer or the winter, or $\ldots$) or the lag $\ell$. Our reason for doing so is that we expect that the major trends and seasonal effects are captured  by the exogenously provided forecasts which are usually constructed from a large number of meteorological variables and Bayesian averaging of many forecasting models.
\end{remark}

\subsection{\textbf{The Impact of Fitting Heavy Tail Distributions}}
Here, we discuss the impact of our decision to check for the presence of heavy tails and to possibly fit generalized Pareto distributions to the marginal laws of the deviations. In order to do so, we provide a detailed examination of the smoothness of the Monte Carlo scenarios and the coverage provided by these scenarios.

\begin{figure}[h]
\centerline{
\includegraphics[width=12cm,height=6cm]{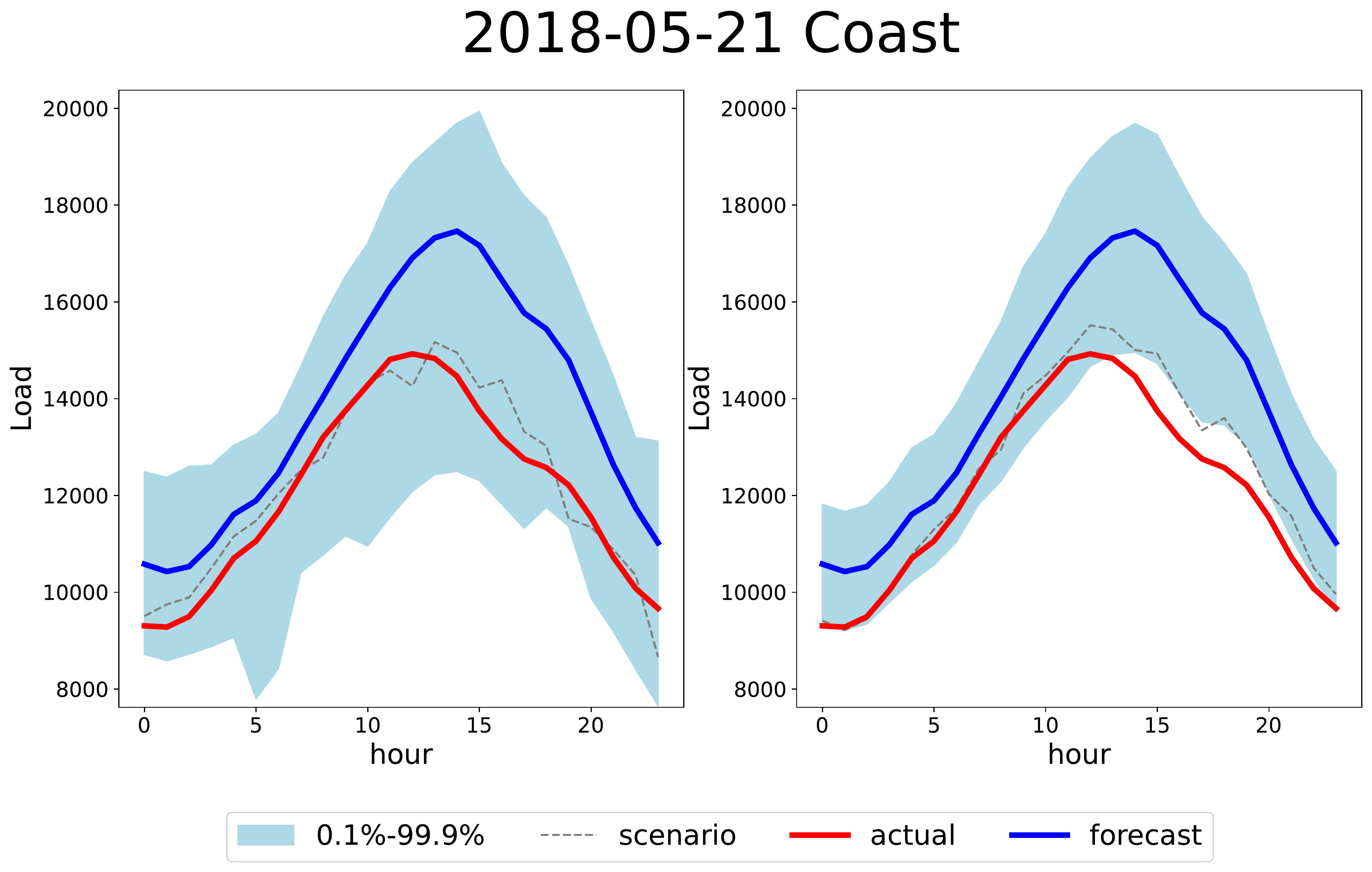}
}
\caption{Actual loads (red curves) and forecasts (blue curve) as well as a gray band covering $10,000$  Monte Carlo scenarios, as well as the sample scenario (dashed lines) with the minimum mean absolute percentage error (MAPE) for the $24$ hours following May 21, 2018. The left pane was produced using the model described in the paper, while the right pane was produced from a Gaussian model ignoring the possible presence of heavy tails in the distributions of the deviations.}
\label{fi:overestimation}
\end{figure}

\vskip 2pt
Repeated overestimation by the forecasts create instances of negative values for the deviations, and if these instances are significant because of their sizes, the empirical distribution of the deviations can exhibit a heavy lower tail which cannot be accounted for by Gaussian distributions.  We illustrate  such an instance in Figure \ref{fi:overestimation} with the example of the load zone Coast on May 21, 2018. The left pane of the figure shows that the actual loads (red curve) hover around the lower boundary of the band created by the Monte Carlo scenarios, especially during the later hours. Recall that these scenarios are obtained by adding sample scenarios of the deviations to the forecasts (blue curve). Fitting a distribution with a heavy lower tail guarantees that the corresponding scenarios will cover the actual values. The latter should be viewed as extremes, possibly rare events, but still in the realm of the model. The right pane shows the same plot produced from Monte Carlo scenarios obtained from a Gaussian model ignoring the presence of heavy tails. While one could think that having a narrower coverage band is desirable, the fact that the actual values are so far from what the model considers to be reasonable is a major flaw of this form of model.

\begin{figure}[H]
\centerline{
\includegraphics[width=12cm,height=6cm]{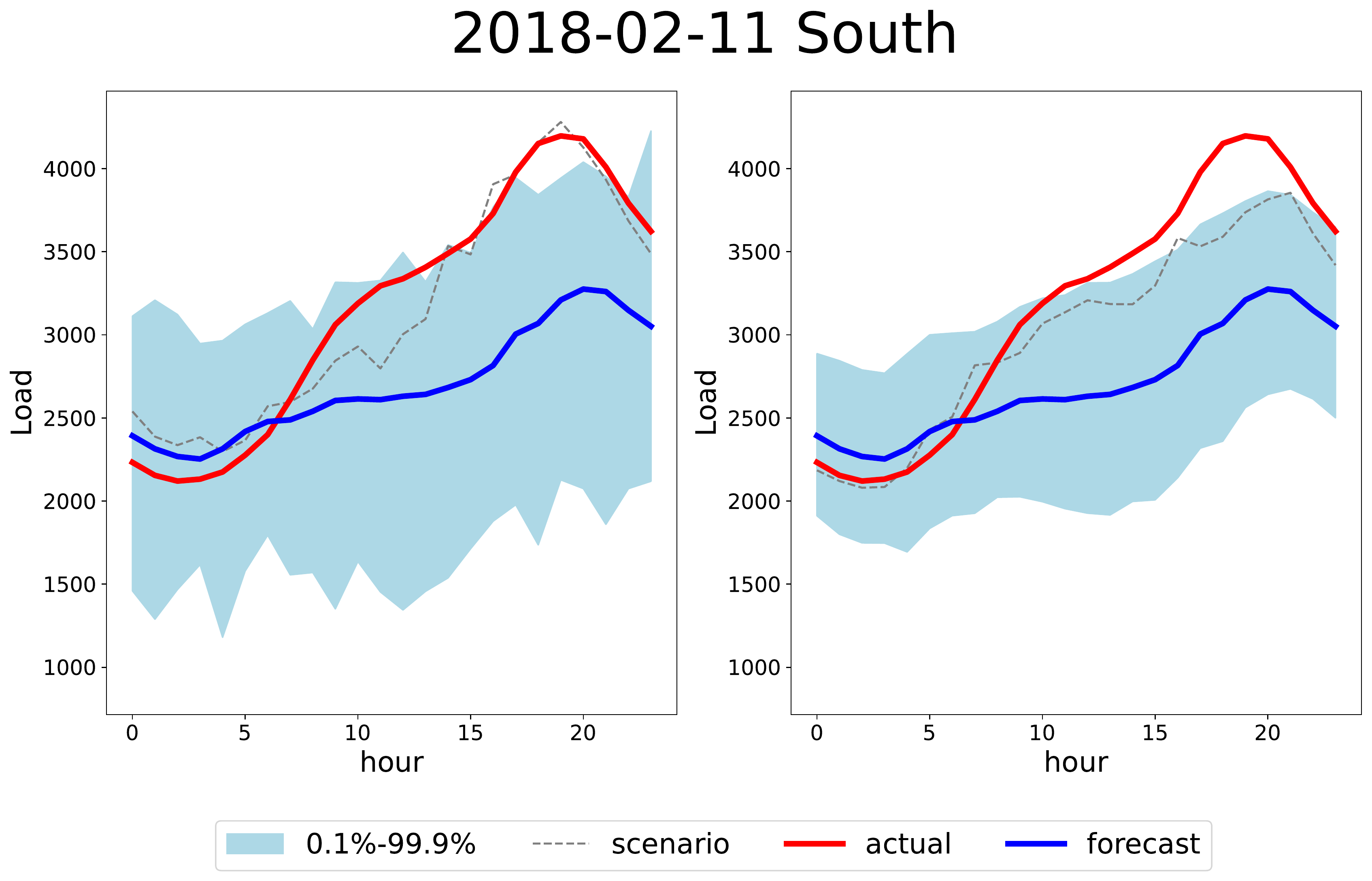}
}
\caption{Same plots as in Figure \ref{fi:overestimation} for February 28, 2018 to illustrate the possible consequences of an underestimation by the forecasts.}
\label{fi:underestimation}
\end{figure}

In a similar fashion, we can discuss the case of underestimation by the forecasts. Clearly such instances can have dire consequences for the management of the electric grid, and they are feared by system operators. Figure \ref{fi:underestimation} provides an illustration which makes a resounding case for the use of heavy tail distributions in the model. We use the example of the load in the West region on February 11, 2018. With the same convention as before, the left pane of Figure \ref{fi:underestimation} shows that the actual loads for the next $24$ hours (red curve), while much higher than the forecasts (blue curve) are mostly covered by the scenarios. So even on the tail end of the possibilities given by the Monte Carlo simulations, it can still be expected to occur with significant probability and the operator of the grid should not dismiss such a pattern.
In contrast, the right pane of Figure \ref{fi:underestimation} shows the result of the same Monte Carlo simulations when the model does not attempt to detect and fit heavy tail distributions. Clearly the curve of actual loads is far beyond the simulation band, and an operator relying on such a model will consider the $24$ actual values as a possibility with zero likelihood.

\vskip 2pt
The two instances discussed above clearly illustrate the pros of checking for the presence of heavy tails in the marginal distributions of the deviations, and when detected, of using generalized Pareto distributions to process the data. 

\section{\textbf{Marginal Model for Wind Power}}
\label{se:wind}

This section will be shorter than the previous one. Indeed, several steps of the analysis are the same as for the loads, so we avoid most repetitions by emphasizing only the significant differences when they are relevant.

\subsection{\textbf{Modeling Steps}}
As before, we introduce the \emph{deviations}:
\begin{center} 
\emph{wind power deviations} = \emph{wind power actuals} – \emph{wind power forecasts}, 
\end{center}
and prepare the data set of wind power deviations exactly as for the loads, the only noticeable difference being that we have now $N_w=264$ production assets instead of $N_z=8$ zones. Obviously, we still have the same $N_{lag}=24$ lags.

\begin{figure}[H]
\centerline{
\includegraphics[width=9cm,height=8cm]{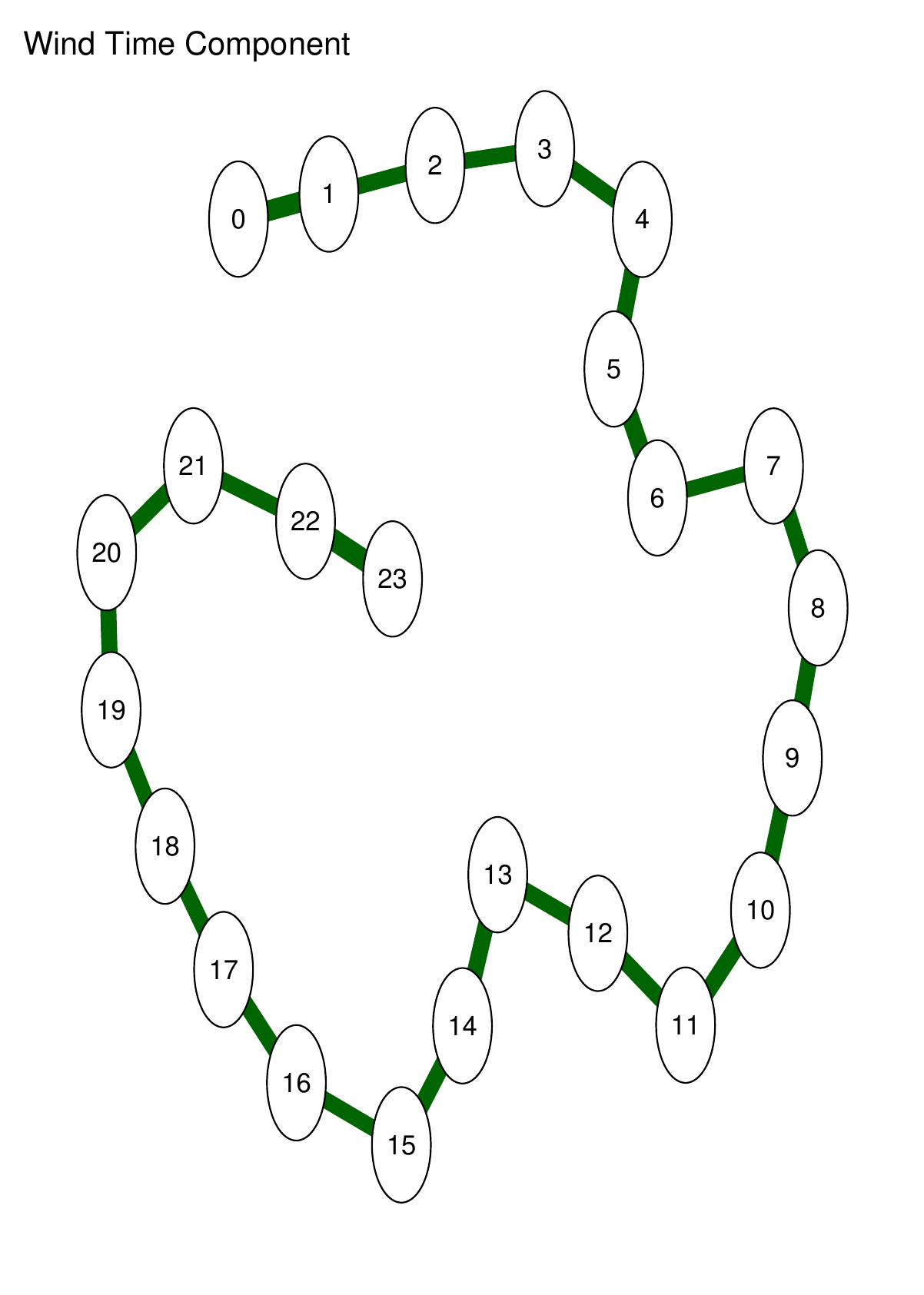}
\hskip 6pt
\includegraphics[width=9cm,height=9cm]{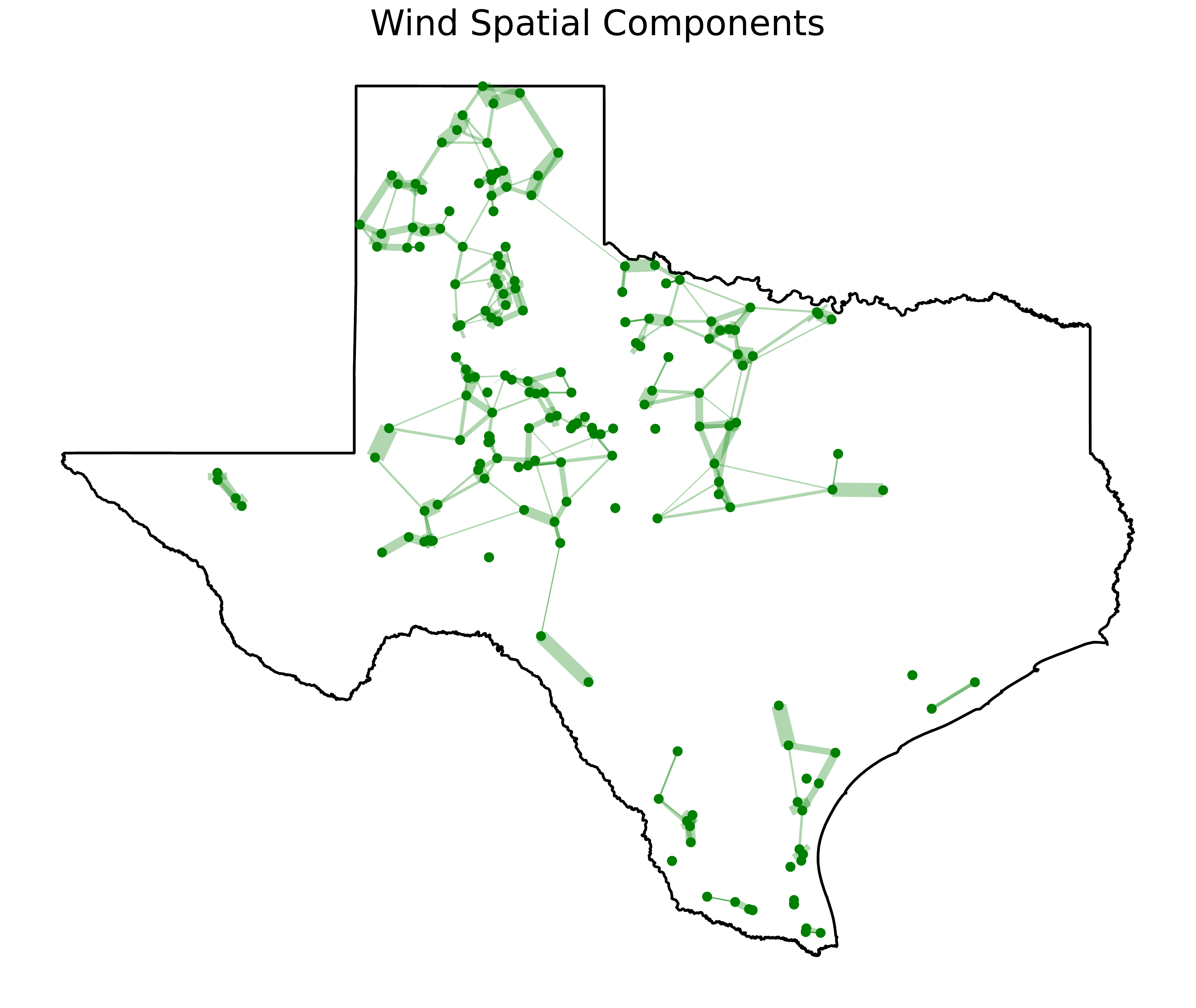}
}
\caption{Temporal component of the conditional correlation (left), spatial component of the conditional correlation (right) using the plain average of $51$ wind ECMWF forecast.}
\label{fi:wind_glasso_ecmwf}
\end{figure}

\vskip 4pt
As for the step 2) above, the marginal distributions do not exhibit heavy tails. Indeed, the power produced by a single wind station is a non-negative number bounded from above by the capacity of the wind farm. So for a specific wind farm, the deviation should have a distribution with support contained in an interval $[-c,+c]$ where $c$ is the capacity of the generation asset. So for each of the $6336$ $(w,\ell)$ asset-lag couples, we use the empirical cumulative distribution function (cdf) to fit  the  marginal distribution of the \emph{stationary} time series, and we apply this cdf to each of the entries to produce a series with uniform marginals, and eventually standard Gaussian after applying the quantile function of the Gaussian distribution. This takes care of the steps 3) - 5) of the strategy used for the load data. Fitting a Gaussian graphical model like we did in step 6) above with the {\tt GEMINI} algorithm to capture the precision and covariance matrices leads to the analog of Figure \ref{fi:load_glasso} which we reproduce as Figure \ref{fi:wind_glasso_ecmwf}.  In the estimation of the spatial components of the covariance matrix using LASSO, we chose the regularization parameter to be a $N_w$-by-$N_w$ symmetric matrix so that the $ij$-th element is proportional to the Euclidean distance between the $i$-th and $j$-th wind farms.

\subsection{Impact of the Quality of the Forecasts.}
\label{sub:forecast_quality}

\begin{figure}[H]
\centerline{
\includegraphics[width=9cm,height=7cm]{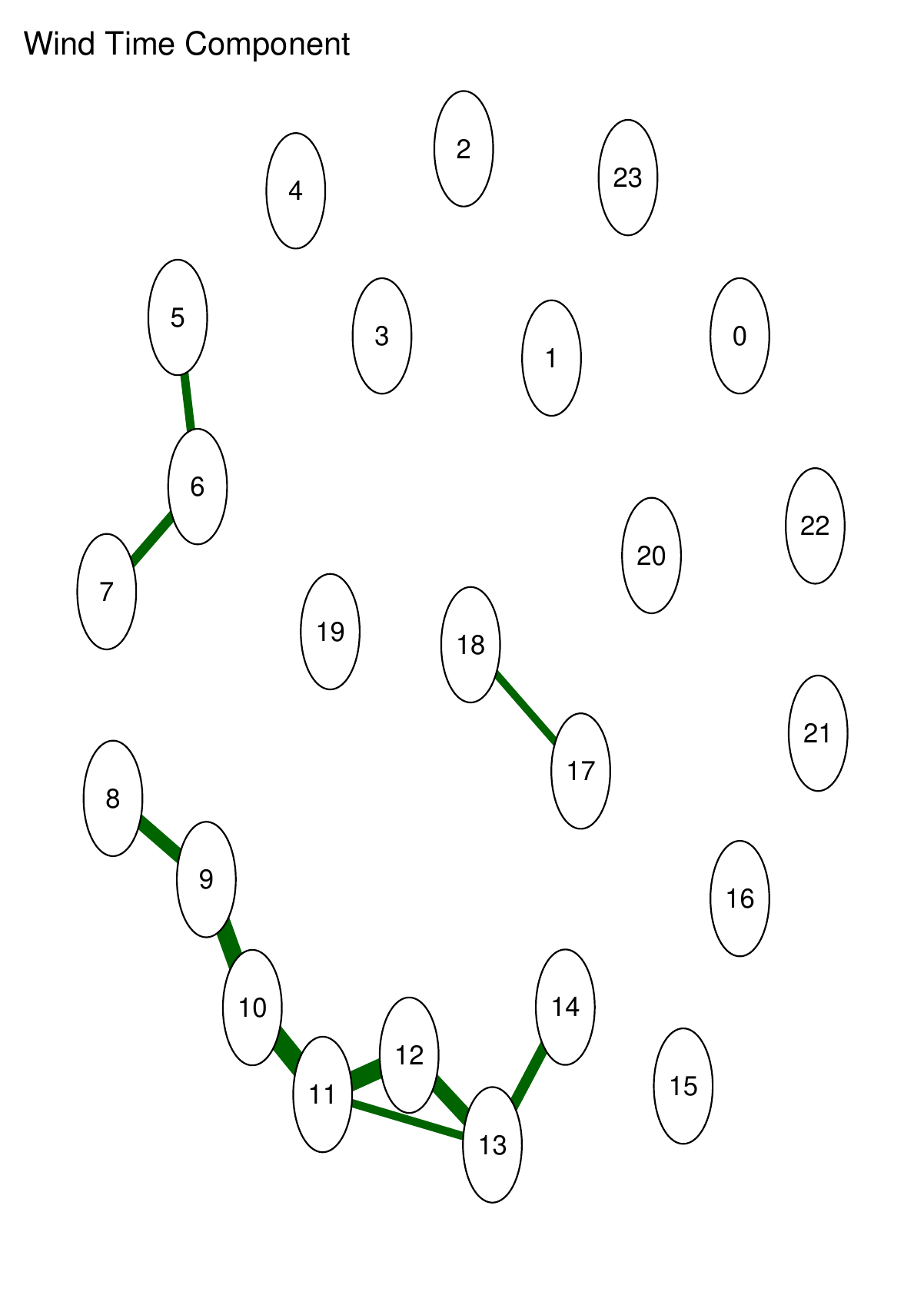}
\hskip 6pt
\includegraphics[width=9cm,height=7cm]{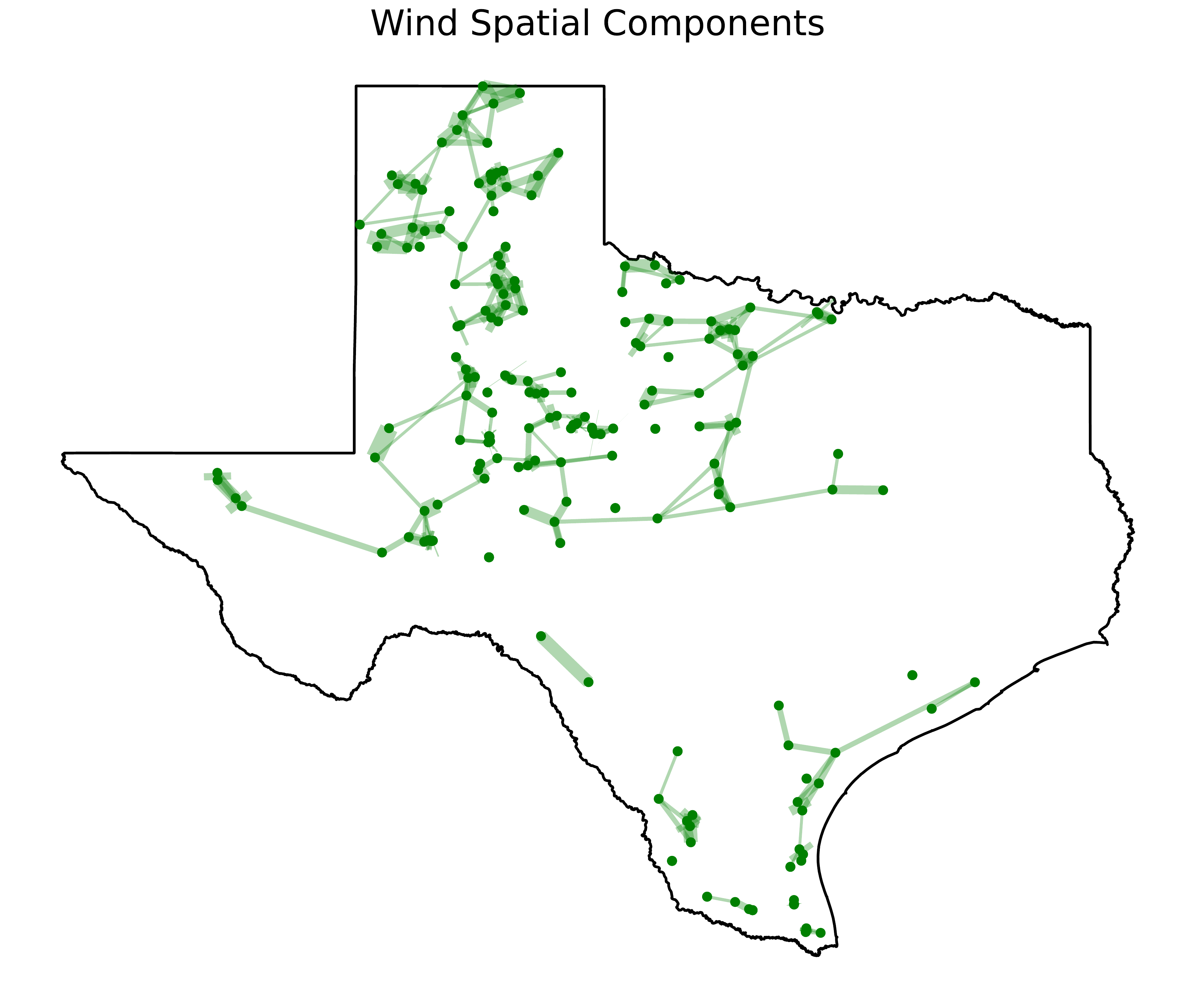}
}
\caption{Temporal component of the conditional correlation (left), spatial component of the conditional correlation (right) using the median of the probabilistic forecast.}
\label{fi:wind_glasso_pf}
\end{figure}

As mentioned in Section \ref{se:data}, we implemented simulation models for wind using two different sets of forecasts: the median of the probability forecasts and a plain average of the $51$ ECMWF forecasts. We show the temporal and spatial components of the conditional correlations using the two forecasts in Figures \ref{fi:wind_glasso_ecmwf} and \ref{fi:wind_glasso_pf}. For both cases, the spatial dependency graphs are consistent with the geographical locations of the
wind power production assets. Figure \ref{fi:wind_glasso_ecmwf} shows that using point forecasts obtained from the ECMWF forecasts, the graphical model shows the usual \emph{Markovian-like} temporal dependence structure given by the clear chain pattern of the temporal dependency graph. However, the temporal dependency graph derived from the probabilistic forecasts doesn't indicate a clear pattern. Moreover, it depends strongly on the choice of the LASSO regularization parameters. For example, the graph shown in Figure \ref{fi:wind_glasso_pf} exhibits many isolated nodes. Moreover, although the temporal dependency graph contains a chain pattern for the hours from $8$ to $14$, the largest correlation coefficient in the graphical model is less than $20\%$.

\subsection{\textbf{Monte Carlo Generation Specifics}}

\begin{figure}[H]
\centerline{
\includegraphics[width=5.5cm,height=5.5cm]{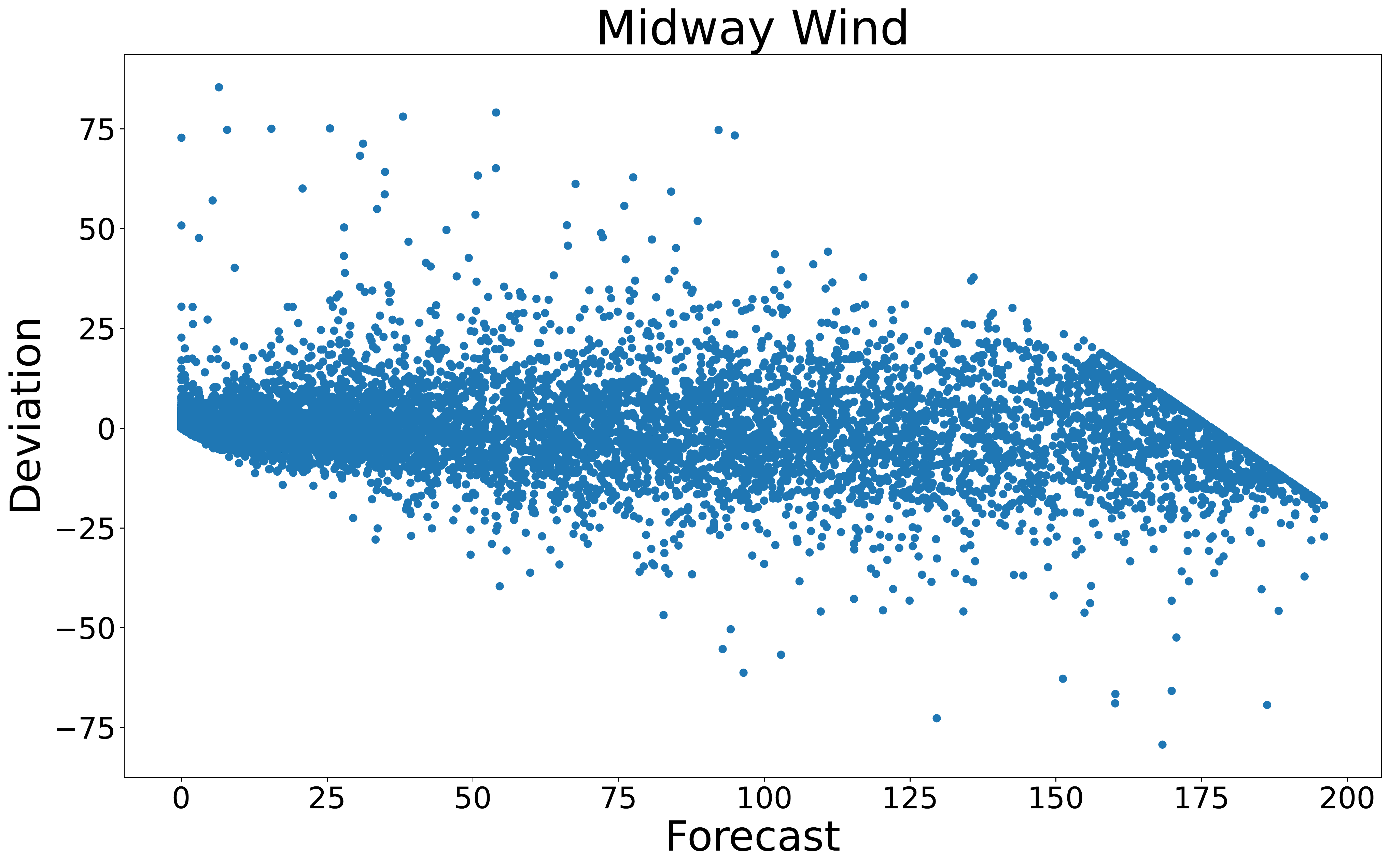}
\hskip 6pt
\includegraphics[width=5.5cm,height=5.5cm]{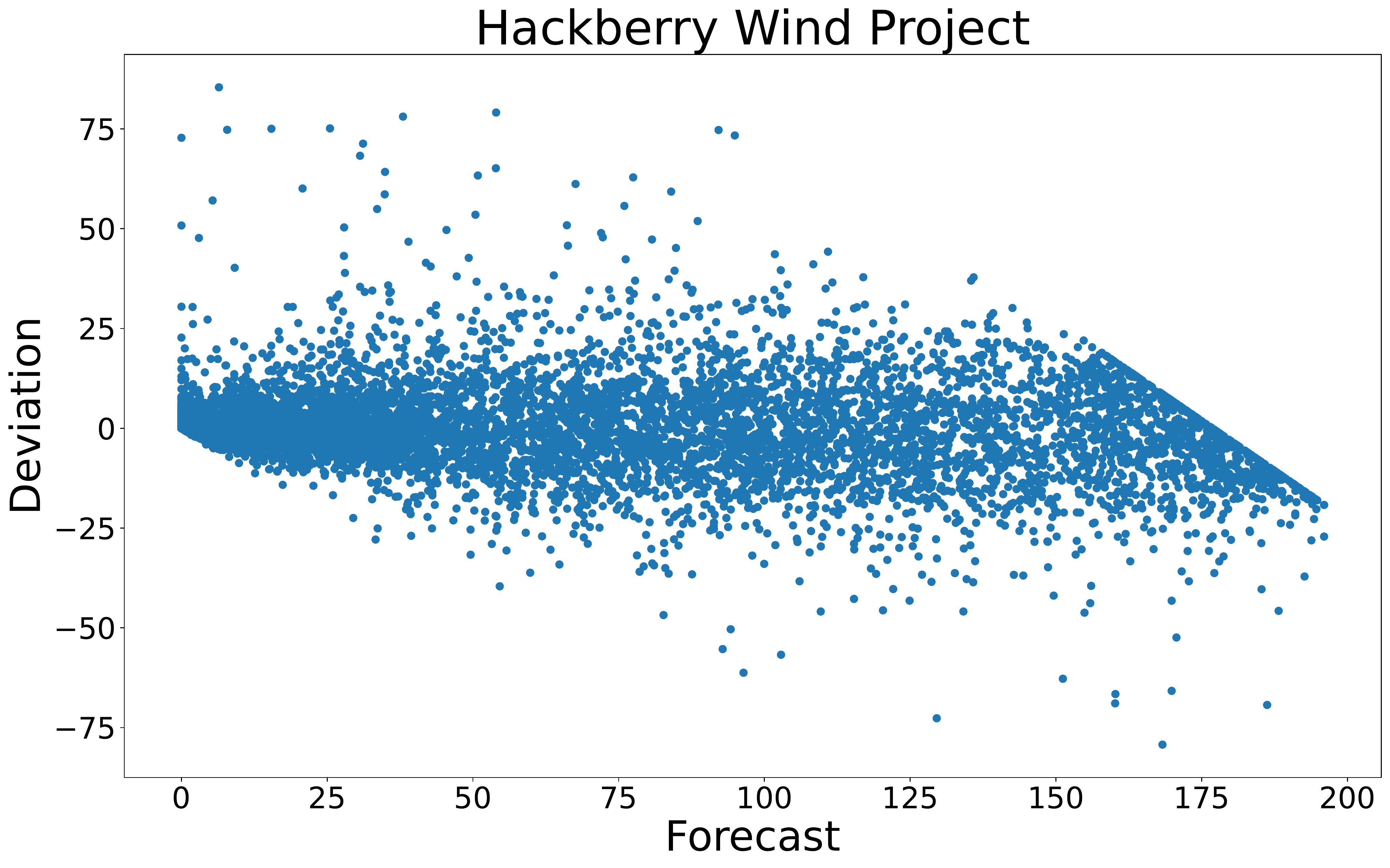}
}
\caption{Scatterplots of the deviations (vertical axis) against the corresponding forecasts (horizontal axis).}
\label{fi:forecast_power}
\end{figure}

Exactly as in the case of the load simulations, we can generate Monte Carlo scenarios for the Gaussian vectors whose distributions were identified from the graphical Gaussian model, and remove the Gaussian marginals by computing the Gaussian cumulative distribution function on the entries of these vectors. This way of building dependencies in the Monte Carlo simulation is reminiscent of the use of the Gaussian copula. However, transitioning to the actual power productions (recall step 7 of our discussion of the load) requires a modicum of care. Indeed, in the case of wind power, the marginal distribution of the deviations strongly depends upon the actual value of the forecast. See Figure \ref{fi:forecast_power} for empirical evidence. 

So in order to account for this phenomenon, we restore the marginal distributions of the scenarios by using the quantile function of conditional distribution of the deviations given a forecast value. We estimate this conditional distribution by aggregating all the couples (forecast, deviation) in the data, grouping the values of the forecasts in a finite set of bins, and estimating the distribution of the actual generation above each bin.  Once this is done, for each couple (asset, lag), we generate the value of each scenario by adding to the forecast, the value of the conditional quantile function for the bin in which the forecast landed on, computed on the Monte Carlo value obtained by the Gaussian simulation made uniform. 

\begin{figure}[H]
\centerline{
\includegraphics[width=16cm,height=8cm]{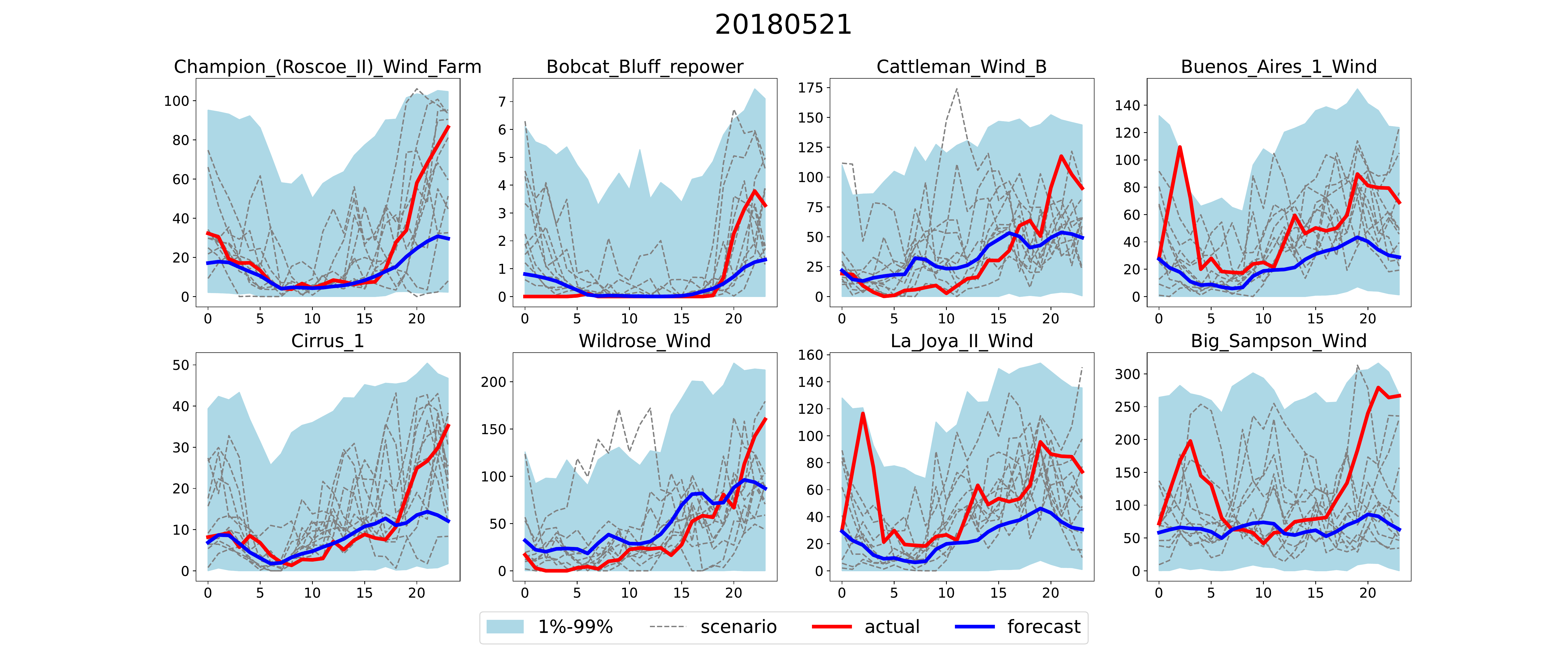}
}
\caption{$1000$ Monte Carlo scenarios generated for the next 24 hours for May 21, 2018, using the mean of the ECMWF forecasts as point forecasts. The red line gives the actual wind power observed over these $24$ hours, the blue line shows the point forecasts for these $24$ hours, while the dashed lines are 10 randomly selected scenarios. The light blue bands are the traces  left by the Monte Carlo scenarios after we remove the smaller $1\%$ and the larger $1\%$ of the bunch.}
\label{fi:wind_scenario_samples_ecmwf}
\end{figure}

\begin{figure}[H]
\centerline{
\includegraphics[width=16cm,height=8cm]{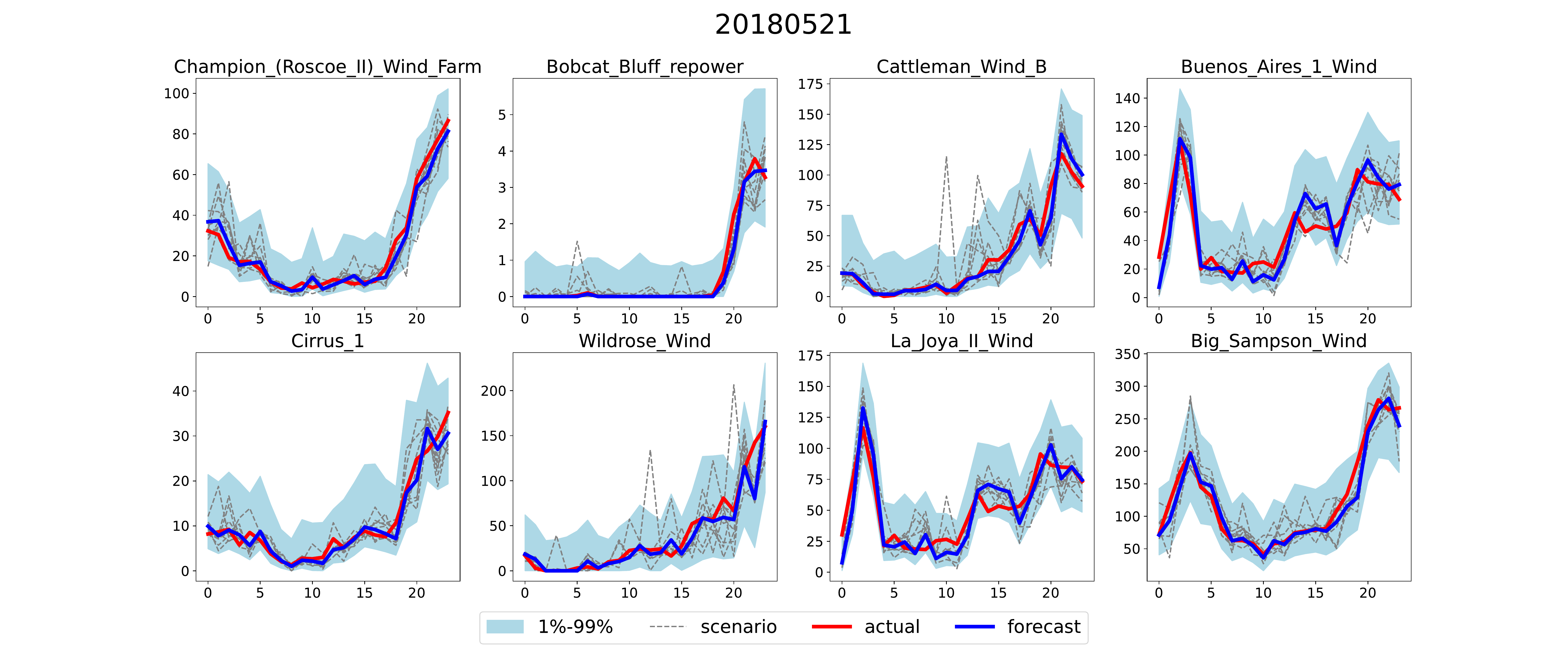}
}
\caption{Same caption as for Figure \ref{fi:wind_scenario_samples_ecmwf} except for the fact that the scenarios are now generated from a model using the median of the probabilistic forecasts as point forecasts.}
\label{fi:wind_scenario_samples_pf}
\end{figure}

\vskip 2pt
For illustration purposes, we show the impact at the asset-level of the quality of the forecasts by the plots of wind power scenarios for $8$ wind farms randomly selected from the $264$ wind farms. The Monte Carlo scenarios are generated from our model using the mean of the ECMWF forecasts in Figure \ref{fi:wind_scenario_samples_ecmwf}, and the median of the probabilistic forecasts in Figure \ref{fi:wind_scenario_samples_pf}. Because the probabilistic forecasts were constructed from a bias correction procedure applied to the ECMWF point forecasts, the forecasts seen in Figure \ref{fi:wind_scenario_samples_pf} are much closer to the actuals than in the case of Figure \ref{fi:wind_scenario_samples_ecmwf}.
The precision with which these forecasts predict the actual values of the wind generation is remarkable.
This being also true for the historical data used to fit the models, the bands covered by the scenarios in Figure \ref{fi:wind_scenario_samples_pf} are narrower.

\section{\textbf{Marginal Model for Solar Power}}
\label{se:solar}

The fact that solar power production follows a very different pattern than wind power production, is in no small part due to the stark differences between nocturnal and diurnal productions, so we adopt a different strategy to fit a model to the data.

In line with our analysis of load demand and wind power production, we work with the deviations between the solar power production actuals and forecasts, i.e., 

\begin{center} 
\emph{solar power deviations} = \emph{solar power actuals} – \emph{solar power forecasts}.
\end{center}
The dataset consists $N_s=226$ production assets and $N_{lag}=24$-hour time horizon for the period 2017-01-01 to 2018-12-31. For each of the $5424$ $(s, l)$ asset-lag couples, the data is in the form of a time series of length $730$. 

In order to make sure that our model (and the subsequent Monte Carlo simulations) are non-anticipative (i.e. do not use a crystal ball), for each given day $t$ of 2018, we collect the historical data comprising the $n$ days preceding $t$, and the days of 2017 which fall in the $2n+1$ days window centered around day $t$ in 2017. We use this approach for any day $t$ in 2018, and rely on the historical data identified in this way to fit the model from which we generate Monte Carlo scenarios for day $t$. In typical experiments we use the default value $n=50$. This is reasonable, even when $t$ is early in the year. For instance, if we are interested in creating scenarios for 2018-07-01, the historical data points to fit the model would be 2017-05-12 to 2017-08-20 and from 2018-05-12 to 2018-06-30.

To transform the data so that the marginal distributions become Gaussian, we first use the empirical cumulative distribution function to produce time series with uniform marginals, and then the standard Gaussian quantile function to end up with Gaussian marginal distributions.

\begin{figure}[H]
\centerline{
\includegraphics[width=9cm,height=8cm]{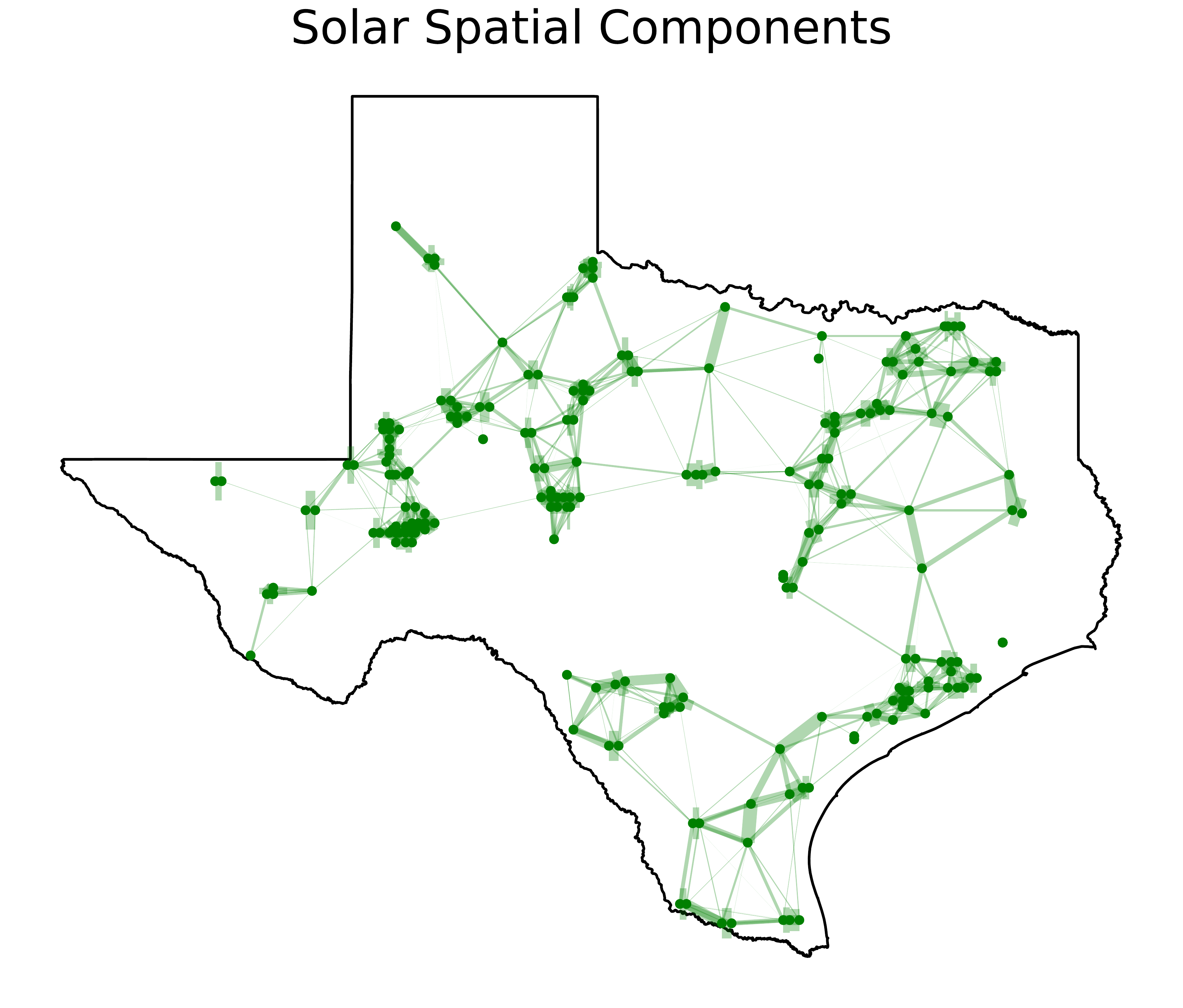}
}
\caption{Spatial component of the conditional correlation for the analysis of the solar deviations using PCA.}
\label{fi:solar_glasso}
\end{figure}

Next, we perform a Principal Component Analysis (PCA) of the transformed data. Denoting by $N$ the number of days used to fit the model, we aggregate the transformed data over all historical days and the assets and end up with a $226N\times24$ matrix, recall that we have $226$ solar stations, one column for each hour of the day, and one row for each couple (asset, day). We then compute the PCA orthonormal basis of $24$ loadings, and a $226N\times24$ matrix of components of the original data onto the loadings basis. Given the proportions of the variation in the data explained by the successive loadings, we chose to keep, for each of the $226N$
couples (asset, day), only the first $k$ entries. The switch from the original data to their first $k$ components is intended to resolve the issue of the diurnal zero production, more than for its \emph{data compression} effect. We now reshape the data into $k\times24$ time series of length $N$, and standardize the data to have zero mean and unit variance. We then fit a graphical Gaussian model using GEMINI to obtain a $226$-by-$226$ and a $k$-by-$k$ covariance matrices that capture the dependencies between the spatial components and  dependencies between the PCA loadings. To generate scenarios, we first draw Monte Carlo samples from the mean zero Gaussian distribution with covariance matrix defined as the tensor product of the two covariance matrices computed from the graphical model.  We then reconstruct $24$-dimensional vectors using the orthonormal basis of the PCA loadings. We complete the simulation by adding back the forecasts. In Figure \ref{fi:solar_glasso}, we show the graph of the spatial conditional temporal dependence, which is again consistent with the geographical locations of the solar power production assets. 

\begin{figure}[H]
\centerline{
\includegraphics[width=18cm,height=7.5cm]{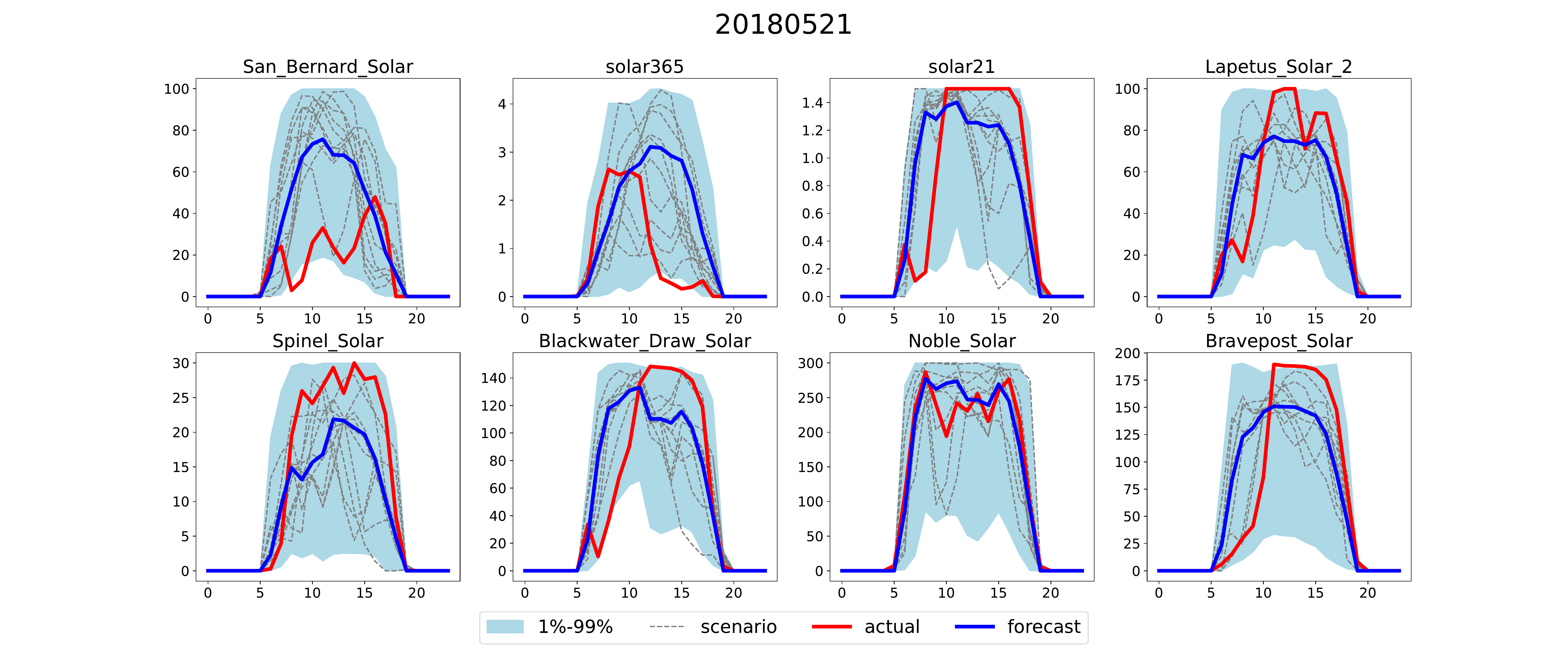}
}
\caption{$1000$ Monte Carlo scenarios generated for the next 24 hours for May 21, 2018, using the plain average of the ECMWF forecasts as point forecasts. The red line gives the actual solar power observed over these $24$ hours, the blue line shows the point forecasts for these $24$ hours, while the dashed lines are $10$ randomly selected scenarios. The light blue bands are the traces left by the Monte Carlo scenarios after we remove the smaller $1\%$ and the larger $1\%$ of the bunch.}
\label{fi:solar_scenario_samples}
\end{figure}

\vskip 2pt
Figure \ref{fi:solar_scenario_samples} shows an example of asset-level solar production scenarios for $8$ solar farms randomly selected from the $226$ solar farms. As we can see, the values of the solar production given by the scenarios are only non-zero during the day. This is the result of our use of the projection of the data on the first $k$ loadings of the PCA.

\section{\textbf{Joint Model and Consolidated Simulation Engine}}
\label{se:altogether}

Once the three marginal models discussed earlier are fitted, and the corresponding simulation engines have been developed, a joint model and a joint simulation engine can be obtained following the steps described in this section.

\subsection{\textbf{In-sample vs Out-of-sample Modeling and Simulation}}
\label{sub:historical}

To guarantee that our model (and the subsequent Monte Carlo simulations) are non-anticipative, for each given day $t$ of 2018, we fit our model to the bunch of historical data identified earlier in the case of solar power generation.
However, if we want to use the wind point forecasts given by the median of the probabilistic forecasts provided by NREL, we do not have historical data for 2017. So when the day $t$ is early in the year, we fit the model with data for days in a window covering $t$, extending in the future by $n$ days and excluding $t$. So, for those values of $t$, the estimation of the model is \emph{in-sample}. This is unfortunate as it should not have to be if we had enough historical data.

\subsection{\textbf{Quantifying and Implementing the Dependencies}}
\label{sub:dependencies}

Load data being available at the zone level only, the strategy for the joint modeling of load, wind power and solar power is based on the following steps.
\begin{enumerate}
\item The first step of the analysis is again a form of \emph{Gaussianizeation} of the historical data of load, wind and solar power productions. We do that separately, just as we did for the marginal models;
\item For each of the load zones and for each lag $\ell$, we aggregate the wind and solar Gaussianized power production historical data from the individual asset levels to the zone levels. So for each of the $8$ load zones and for each lag $\ell$, we sum up all the Gaussianized time series of the individual wind asset deviations to create one single wind power deviation series for this zone. Similarly, for each of the $8$ load zones and for each lag $\ell$, we sum up all the Gaussianized time series of the individual solar asset deviations to create one single solar power deviation series for this zone; 
\item Next, for each of the load zones, we aggregate the zonal load, wind and solar Gaussianized power productions across time. As solar plants only generate power during the day, we only aggregate across those hours. In particular, if we denote the lags during sunrise and sunset as $\ell_{sunrise}$ and $\ell_{sunset}$ respectively, then we sum up all the Gaussianzed zonal load time series to create one single load series for that zone during that day. Similarly, we aggregate wind (and solar) to create one single series for that zone during that day. 

\item For each load zone $z$, we concatenate the three time series into one three dimensional time series and fit a joint Gaussian graphical model at the zone level, and using a plain {\tt glasso} algorithm. For this step, we do not use the {\tt gemini} variation on the {\tt glasso} algorithm because the temporal dependency graphs of load, wind and solar are sufficiently different for us to be convince that the assumption of \emph{joint separability} of the covariance matrix is unreasonable.
\item When running the {\tt glasso} algorithm with load. solar power and wind power, we find that the nodes of the dependency graph corresponding to wind are disconnected from the nodes corresponding to load and solar. Since we are still in the \emph{Gaussian world} when we perform this step of the analysis, our conclusion is that, at least at the zonal level, wind power generation is statistically independent of the electricity demand and the solar power generation. As a.result, wind is modelled separately and Monte Carlo scenarios are generated for wind power from the marginal model described in Section \ref{se:wind}
\item As for diurnal load and solar power, Figure \ref{fi:load_solar_joint} provides the spatial dependency graph as given by the precision matrix,
while Figure \ref{fi:load_solar_joint_heatmap} provides a heat-map for the $16\times 16$ correlation matrix of the diurnal aggregates of load and solar power over the $8$ load zones.
The dependence highlighted in Figure \ref{fi:load_solar_joint} by the links between the load zones is consistent with the geographical locations of these zone, and the fact that the node corresponding to load in the "Far-West" zone is isolated is not a surprise. Links between different solar zones are also natural. Strong links between load and solar power for the same zone are also significant and expected, for example load and solar in zone South Central, or zones East and Coast, and to a lesser degree zone South.

\begin{figure}[H]
\centerline{
\includegraphics[width=16cm,height=10cm]{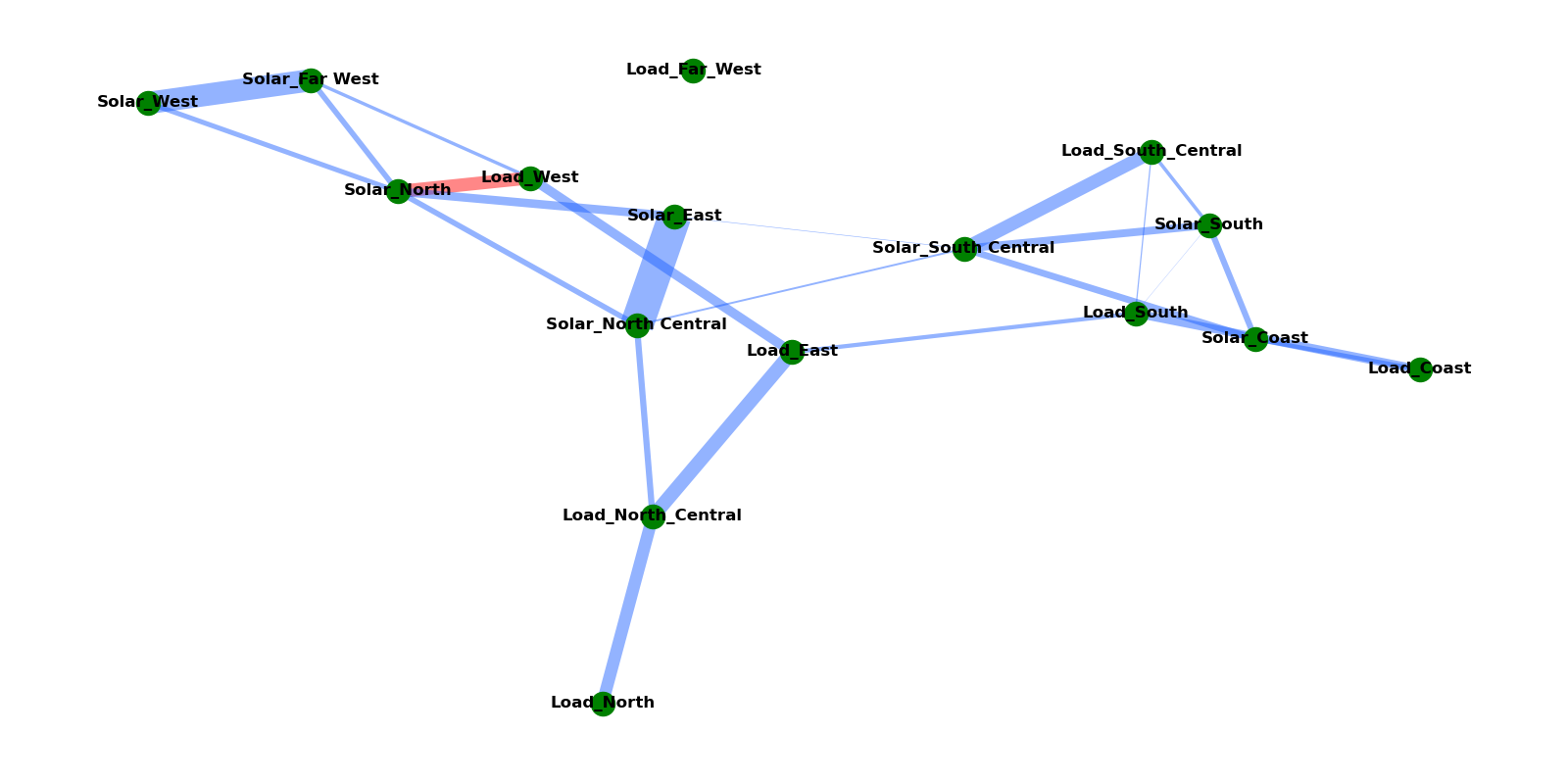}
}
\caption{Graphical LASSO estimate of the conditional correlation of load and solar power in the joint model described in the text.}
\label{fi:load_solar_joint}
\end{figure}

As expected, the corresponding correlation matrix whose heat-map is given in Figure \ref{fi:load_solar_joint_heatmap} is not sparse. The bottom-right fourth of the matrix confirms the strong correlations between the zone level diurnal solar power productions. These correlations are weaker among the zonal loads. The green blocks appearing in the bottom-left quadrant of the matrix confirm the remarks we made above concerning the correlations between load and power production in some regions.

\begin{figure}[H]
\centerline{
\includegraphics[width=22cm,height=12cm]{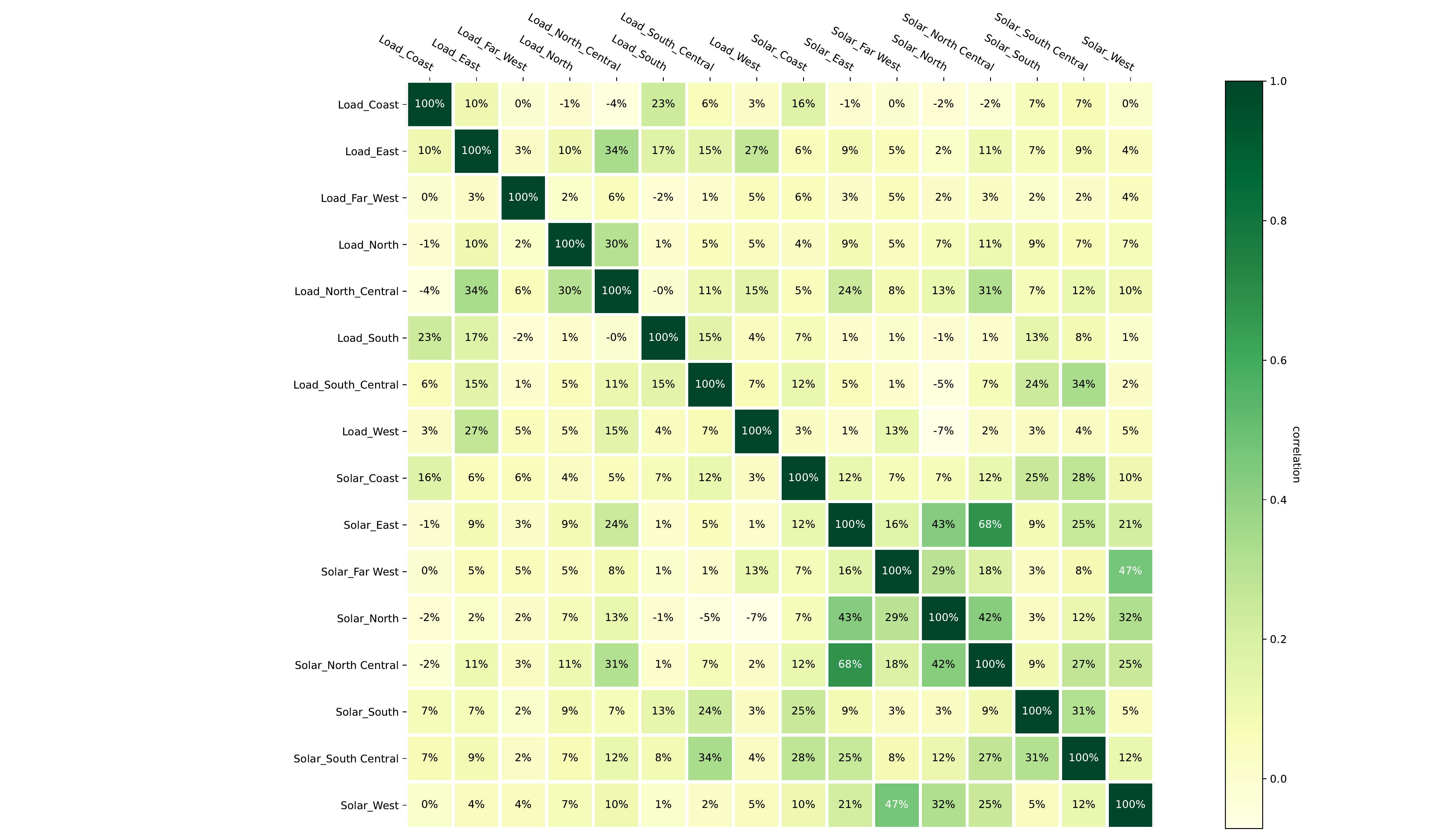}
}
\caption{The correlation heatmap of the load-solar joint model.}
\label{fi:load_solar_joint_heatmap}
\end{figure}

\end{enumerate}

\subsection{\textbf{Putting Everything Together with Conditional Simulation}}

Giving the above discussion, in order to generate Monte Carlo scenarios for the entire system on a given day $t$, we can first simulate samples of $24$ hours wind power generation for the $264$ wind farms using the marginal model introduced and fitted in Section \ref{se:wind}. Next, and independently of the scenarios generated for wind power, we generate jointly correlated Monte Caro scenarios for load and solar power according to the following steps based on conditional simulation of Gaussian processes.
\begin{enumerate}
\item We generate jointly Gaussian scenarios for load and zonal solar power as their daily sum of the lags from $\ell_{sunrise}$ to $\ell_{sunset}$. We do just that using the joint Gaussian model estimated from the {\tt glasso} algorithm implemented as described above.
\item Next for each such scenario, we generate a $8\times 24$ ($8$ for each load zone and $24$ for the $24$ time lags) Gaussian load Monte Carlo scenario from the marginal distribution fitted using the {\tt gemini} algorithm as described in Section \ref{se:load}, conditional on the fact that the sum of the lags in the range $[\ell_{sunrise}, \ell_{sunset}]$ are exactly equal to the values of the load scenario generated in step 1. We do that for the eight load zones \emph{simultaneously} to make sure that we do not wipe out the dependencies between the zones!
\item Next for each scenario generated in step 1, we generate a $226\times 24$ ($226$ for each of the solar assets, and $24$ for the $24$ time lags) Gaussian load Monte Carlo scenario from the marginal distribution fitted using the PCA based algorithm described in Section \ref{se:solar}, conditional on the fact that the zonal aggregates of the lags in the range $[\ell_{sunrise}, \ell_{sunset}]$ are exactly equal to the values of the particular solar scenario generated in step 1. We do that for the $226$ solar assets \emph{simultaneously} to make sure that we do not wipe out the dependencies between the individual asset!
\item Once we have Monte Carlo scenarios for the Gaussian versions of zonal loads and individual solar assets, we can undo the Gaussian nature of the scenarios as explained in Section \ref{se:load} and Section \ref{se:solar} and reinstate the non-necessarily Gaussian marginal distributions. 
\end{enumerate}

\section{\textbf{Conclusion}}
\label{se:conclusion}
In this paper, we propose an original model for electricity demand and wind and solar power generation together with an implementation producing joint Monte Carlo simulations for the purpose of unit commitment and economic dispatch performed by highly sophisticated optimization programs. For this, while electricity load can remain at the zonal level, wind and solar productions need to be modeled and simulated at the individual asset level. The main thrust of our model is to capture the possible and intricate dependencies both temporal and geographical. We consider hourly data, but our model and our simulation algorithms are agnostic to the time frequency and can be ported to higher frequencies as long as the historical data is available.

The main novelty of our approach resides on 1) checking for the presence of heavy tails in the marginal distributions and the use of  generalized Pareto distributions to \emph{Gaussianize} the data in the spirit of the use of Gaussian copulas in portfolio theory; 2) using graphical Gaussian model and producing precision matrix proxies with the {\tt gemini} variation on the {\tt glasso} algorithm when appropriate, disentangling the temporal and spatial contributions to the correlations, and allowing for implementations with high dimensional state variables. We illustrate the  potential of the model and its implementation on synthetic data produced by NREL. Our implementation, called PGscen, is part of the GitHub \begin{center}
https://github.com/PrincetonUniversity/PGscen
\end{center}

\section{\textbf{Acknowledgments}}
Both authors were partially supported by ARPA-E grants DE-AR0001289 and DE-AR0001390 under the PERFORM program of the US Department of Energy. A preliminary version of the paper was presented in November 2020 during the INFORMS Annual Meeting, and on December 21, 2020 at a DOE Review Meeting. We would like to thank Mike Ludkovski (University of California at Santa Barbara), Ronnie Sircar (Princeton University) and Glen Swindle (Scoville Risk Partners) for enlightening conversations on the content of the paper, and Michal Grzadkowski for the development of the GitHub package PGscen implementing the model presented in the paper.

\bibliographystyle{plain}
 \small
\bibliography{PERFORM.bib}

\end{document}